\documentclass[twocolumn]{aastex61}
\pdfoutput=1
\usepackage{amsmath}
\usepackage{url}

\newcommand{\msun} {M_\odot}
\newcommand{\lsun} {L_\odot}
\newcommand{\Teff} {T_{\rm eff}}
\newcommand\gta{\lower 0.5ex\hbox{$\ \buildrel > \over \sim\ $}}
\newcommand\lta{\lower 0.5ex\hbox{$\ \buildrel < \over \sim\ $}}

\begin{document}

\title{On the Spectral Evolution of Hot White Dwarf Stars. \\ III. The PG 1159$-$DO$-$DB$-$DQ Evolutionary Channel Revisited}

\author{A. B\'edard}
\affiliation{D\'epartement de Physique, Universit\'e de Montr\'eal, Montr\'eal, QC H3C 3J7, Canada; email: antoine.bedard@umontreal.ca \vspace*{6mm}}

\author{P. Bergeron}
\affiliation{D\'epartement de Physique, Universit\'e de Montr\'eal, Montr\'eal, QC H3C 3J7, Canada; email: antoine.bedard@umontreal.ca \vspace*{6mm}}

\author{P. Brassard}
\affiliation{D\'epartement de Physique, Universit\'e de Montr\'eal, Montr\'eal, QC H3C 3J7, Canada; email: antoine.bedard@umontreal.ca \vspace*{6mm}}

\shorttitle{The PG 1159$-$DO$-$DB$-$DQ Evolution}
\shortauthors{B\'edard, Bergeron \& Brassard}

\begin{abstract}

We continue our comprehensive theoretical investigation of the spectral evolution of white dwarfs based on sophisticated simulations of element transport. In this paper, we focus on the transformation of PG 1159 stars into DO/DB white dwarfs due to the gravitational settling of heavy elements, and then into DQ white dwarfs through the convective dredge-up of carbon. We study the impact of several physical parameters on the evolution of the surface carbon abundance over a wide range of effective temperature. In the hot PG 1159 and DO phases, our calculations confirm that the temperature of the PG 1159-to-DO transition depends sensitively on the stellar mass and the wind mass-loss rate. We show that measured carbon abundances of DOZ white dwarfs are mostly accounted for by our models, with the notable exception of the coolest DOZ stars. In the cooler DB and DQ phases, the predicted atmospheric composition is strongly influenced by the stellar mass, the thickness of the envelope, the initial carbon content, the efficiency of convective overshoot, and the presence of residual hydrogen. We demonstrate that, under reasonable assumptions, our simulations reproduce very well the observed carbon abundance pattern of DQ stars, which thus allows us to constrain the extent of the overshoot region in cool helium-rich white dwarfs. We also argue that our calculations naturally explain a number of recent empirical results, such as the relative excess of low-mass DQ stars and the presence of trace hydrogen and/or carbon at the surface of most DC and DZ stars.

\end{abstract}

\section{Introduction} \label{sec:intro}

It is observationally well established that white dwarf stars come in two main groups: those with hydrogen-dominated atmospheres, and those with helium-dominated atmospheres. In most cases, the surface is chemically pure, essentially because the strong gravitational field makes the process of gravitational settling highly efficient. Nevertheless, several objects unequivocally show traces of elements other than the primary constituent in their outer layers. Furthermore, multiple studies have demonstrated that a significant fraction of white dwarfs experience radical changes of atmospheric composition as they cool, a phenomenon known as spectral evolution. This empirical evidence indicates that various element transport mechanisms, such as atomic diffusion, convective mixing, stellar winds, and matter accretion, counteract the action of gravitational settling in white dwarf envelopes (see \citealt{bedard2020,bedard2022}, hereafter Papers I and II, for comprehensive overviews).

One relatively well-studied case of spectral evolution is that of stars possessing helium-rich, carbon-polluted atmospheres. It is often referred to as the PG 1159$-$DO$-$DB$-$DQ evolutionary channel, after the chronological sequence of observed spectral types. The commonly accepted scenario begins with an extremely hot PG 1159 star whose envelope consists of a mixture of helium, carbon, and oxygen in similar proportions. The carbon and oxygen are initially supported in the outer layers by a weak radiative wind. As the star cools, the wind dies and gravitational settling causes the carbon and oxygen to sink out of sight, thereby producing a DO and then DB white dwarf with a pure-helium atmosphere. Then, a convection zone develops in the helium layer and eventually catches up some of the settling carbon, which is thus brought back to the surface. The result is a helium-dominated, carbon-polluted atmosphere corresponding to the DQ spectral class. Finally, as the convective region reaches a maximum depth, the dredged-up carbon partially sinks back into the star, leading to a decline of the surface carbon abundance with decreasing effective temperature.

Several theoretical studies have quantitatively assessed the different phases of this evolutionary scenario by carrying out time-dependent calculations of element transport in stellar models. The PG 1159-to-DO transition was investigated extensively by \citet{unglaub2000} and \citet{quirion2012}. They showed that the only viable mechanism allowing PG 1159 stars to keep their peculiar atmospheric composition at the beginning of the cooling sequence is the occurrence of weak mass loss competing against gravitational settling above $\Teff \sim 75,000$ K. Furthermore, \citet{dehner1995}, \citet{fontaine2002}, \citet{althaus2004}, and \citet{althaus2005a} performed similar computations but focused primarily on the chemical structure in the cooler DB phase ($\Teff \sim 20,000 - 30,000$ K). They demonstrated that the resulting DB white dwarf is characterized by a double-layered envelope, with a pure-helium layer on top of a mantle retaining the initial composition of the PG 1159 progenitor. In other words, even at low effective temperature, element separation is still going on in the envelope and the state of diffusive equilibrium has not been reached. This is because the diffusion timescales are much longer at the bottom than at the top of the envelope. We note that among the studies quoted above, only those of the La Plata group \citep{althaus2004, althaus2005a} considered the effect of thermal diffusion, which was shown by \citet{althaus2004} to be non-negligible for this particular problem.

The DB-to-DQ transition around $\Teff \sim 10,000$ K has also been the subject of many investigations. The first detailed models of carbon dredge-up were computed by \citet{pelletier1986}, who were able to reproduce very well the trend of decreasing surface carbon abundance with decreasing effective temperature observed among DQ stars (see \citealt{coutu2019} for an up-to-date empirical picture). However, this agreement could only be obtained by assuming envelopes much thinner ($q_{\rm env} = M_{\rm env}/M \sim 10^{-3.5}$) than predicted by the standard theory of stellar evolution ($q_{\rm env} \sim 10^{-2.0}$). Although their work represented a pioneering step forward, it suffered from several shortcomings: (1) they adopted a semi-evolutionary approach in which the feedback of composition changes on the stellar models is ignored, (2) they used initial chemical structures consisting of a single-layered, pure-helium envelope atop a pure-carbon core (that is, they assumed that complete element separation is achieved in previous evolutionary phases), (3) they neglected thermal diffusion and convective overshoot in the transport equations, and (4) they relied on pieces of constitutive physics, especially radiative opacities, that have since been considerably improved.

\citet{macdonald1998} carried out analogous calculations into which they incorporated a few upgrades: they adopted a full evolutionary approach in which element transport is coupled to the cooling process, and they employed modern radiative opacities from the OPAL project \citep{iglesias1996}. They demonstrated that the new opacities result in deeper convection zones, hence making carbon dredge-up possible in thicker helium envelopes. Nevertheless, their computations predicted a monotonic increase of the atmospheric carbon abundance with time, in sharp contrast with the observations, indicating that their models suffered from some unknown issue.

Following the realization by \citet{dehner1995} that DB white dwarfs harbor double-layered envelopes, it became clear that the next generation of DQ models would need to take into account the full chemical evolution starting from the PG 1159 phase. In this new paradigm, the carbon seen at the surface of DQ white dwarfs is dredged up not from the core, as envisioned by \citet{pelletier1986}, but rather from the PG 1159-like layer at the bottom of the envelope. Our group presented in \citet{dufour2005} a set of such evolutionary sequences (see also \citealt{fontaine2005} and \citealt{brassard2007}). These calculations nicely matched the carbon abundance pattern of DQ stars, without the need to invoke very thin envelopes, thanks to the use of the OPAL opacities and the correct treatment of the past chemical history. Similar cooling sequences were published by the La Plata group in \citet{althaus2005a}, \citet{scoccola2006}, and \citet{camisassa2017}. These works incorporated further notable improvements: they computed the full stellar evolution from the main sequence to the white dwarf phase, and they included thermal diffusion and convective overshoot in the element transport scheme. \citet{camisassa2017} additionally considered the effect of using detailed non-grey atmospheres as boundary conditions for the stellar models. Although these calculations are in principle the most sophisticated available, they appear to be affected by a problem akin to that seen in \citet{macdonald1998}: they fail to reproduce the empirical trend of decreasing surface carbon abundance at low effective temperature.

Finally, it is important to point out that all aforementioned studies ignored non-ideal effects in the modeling of atomic diffusion \citep{beznogov2013}, which were recently shown by \citet{koester2020} to be significant at the bottom of cool DQ envelopes. However, we note that the quantitative results reported by \citet{koester2020} are flawed because they are based on single-layered envelope models in diffusive equilibrium, an assumption known to be invalid since the work of \citet{dehner1995}. So far, full evolutionary calculations incorporating a non-ideal treatment of diffusion have only been performed for hydrogen-rich DA white dwarfs by \citet{althaus2020b}.

In Paper II, we presented and discussed a state-of-the-art simulation of the PG 1159$-$DO$-$DB$-$DQ evolutionary channel that circumvents all of the shortcomings highlighted above. We also demonstrated that our evolutionary sequence, which assumes standard values of the input physical parameters, is in excellent agreement with the observed carbon abundance pattern of DQ white dwarfs. The goal of the present paper is to introduce many more analogous calculations in order to explore the model parameter space relevant to the PG 1159-to-DO and DB-to-DQ transitions in detail. Although this has been done to some extent before, most notably in \citet{pelletier1986}, the theoretical advances achieved since then definitely warrant a new comprehensive investigation. In Section \ref{sec:comp}, we succinctly describe the physical and numerical setup of our simulations. Section \ref{sec:res} presents our results regarding the influence of various stellar properties (such as the total mass, the thickness of the envelope, the strength of the wind, and the extent of convective overshoot) on the chemical evolution. We then discuss the astrophysical implications of our findings in Section \ref{sec:disc}. Finally, our conclusions are summarized in Section \ref{sec:conclu}.

\vspace{4mm}

\section{Computations} \label{sec:comp}

We produce time-dependent simulations of element transport in cooling white dwarfs using the STELUM evolutionary code, which is described at length in Paper II. Briefly, STELUM employs a finite-element scheme to build and evolve complete stellar models, down from the center up to the very surface. Furthermore, it solves the equations governing the transport of chemical species along with the equations of stellar structure in a self-consistent way. Atomic diffusion, macroscopic mixing, and mass loss or accretion are treated as time-dependent diffusive processes and can all be considered simultaneously. Thanks to these features, STELUM is a powerful tool to study the spectral evolution of white dwarfs from a theoretical perspective. We refer the reader to Paper II for details regarding the constitutive physics and numerical techniques of the code.

In the present work, we start our calculations from very hot white dwarf models ($\Teff = 90,000$ K) consisting of a homogeneous carbon/oxygen core surrounded by a homogeneous helium/carbon/oxygen envelope typical of PG 1159 stars. Unless otherwise stated, our evolutionary sequences include chemical diffusion, gravitational settling, and thermal diffusion through the formalism of \citet{burgers1969}, and take non-ideal effects into account following \citet{beznogov2013}. In the rest of the paper, the non-ideal treatment of microscopic diffusion is simply referred to as non-ideal diffusion for conciseness. We consider standard convective mixing according to the mixing-length theory, as well as mixing due to convective overshoot using the prescription of \citet{freytag1996} for the overshoot diffusion coefficient,
\begin{equation}
D_{\rm ov} = D_0 \ \exp{ \left( \frac{-2 |r-r_0|}{f_{\rm ov} \, H_0} \right) } .
\label{eq:dov}
\end{equation}
In this expression, $r$ is the radial coordinate, $r_0$ is the radius of the convective boundary, $D_0$ and $H_0$ are the convective diffusion coefficient and pressure scale height at $r_0$, and $f_{\rm ov}$ is a numerical factor controlling the extent of the overshoot region\footnote{Note that in the overshoot region, the energy transport is assumed to be purely radiative, meaning that we consider only the particle transport (and not the energy transport) arising from convective overshoot.}. Equation \ref{eq:dov} is applied both at the upper and lower boundaries of the convection zone, but it is mainly the lower overshoot region that is relevant for this paper. We also include an outward wind in the early evolution using by default the formula of \citet{blocker1995} for the mass-loss rate,
\begin{equation}
\dot{M} = -1.29 \times 10^{-15} \left( \frac{L}{\lsun} \right)^{1.86} \msun \ \rm{yr}^{-1} ,
\label{eq:mdot}
\end{equation}
where $L$ is the surface luminosity. This wind model is admittedly very crude; in particular, it neglects the metallicity dependence predicted by radiative wind theory \citep{unglaub2000}. It is nevertheless commonly employed due to our poor theoretical understanding of the source of mass loss in hot white dwarfs, and also because it has the important advantage of computational simplicity and flexibility. Finally, residual nuclear burning and radiative levitation are ignored throughout our calculations for simplicity.

As will become apparent below, the variation of the atmospheric composition along the PG 1159$-$DO$-$DB$-$DQ evolutionary channel depends sensitively on a number of model parameters. The main parameters studied in this work are the total mass ($M$), the fractional mass of the envelope ($q_{\rm env}$), the initial carbon mass fraction in the envelope ($X_{\rm C,i}$), the mixing-length parameter ($\alpha$), the overshoot parameter ($f_{\rm ov}$), and the mass-loss rate ($\dot{M}$)\footnote{According to pre-white dwarf evolutionary calculations, some of these parameters (especially $M$, $q_{\rm env}$, and $X_{\rm C,i}$) should be correlated \citep{miller-bertolami2006,althaus2009b}. However, given the historical thin-envelope interpretation of DQ white dwarfs, we prefer here to treat them as independent parameters in order to explore their individual influence on the chemical evolution.}. In Section 3.1 of Paper II, we presented in detail a typical PG 1159$-$DO$-$DB$-$DQ evolutionary sequence, which we henceforth refer to as our standard or reference sequence. For this simulation, we assumed $M = 0.6 \ \msun$, $\log q_{\rm env} = -2.0$, $X_{\rm C,i} = 0.5$ (along with $X_{\rm He,i} = 0.4$ and $X_{\rm O,i} = 0.1$), $\alpha = 1.0$, $f_{\rm ov} = 0.075$, and $\dot{M}$ given by Equation \ref{eq:mdot}. In the following, we present a series of similar calculations in which we vary each of the key parameters one at a time to investigate their individual effect on the chemical evolution.

\section{Results} \label{sec:res}

\subsection{The Reference Sequence} \label{sec:res_ref}

\begin{figure*}
\centering
\includegraphics[width=2.\columnwidth,clip=true,trim=2.0cm 8.25cm 2.0cm 9.5cm]{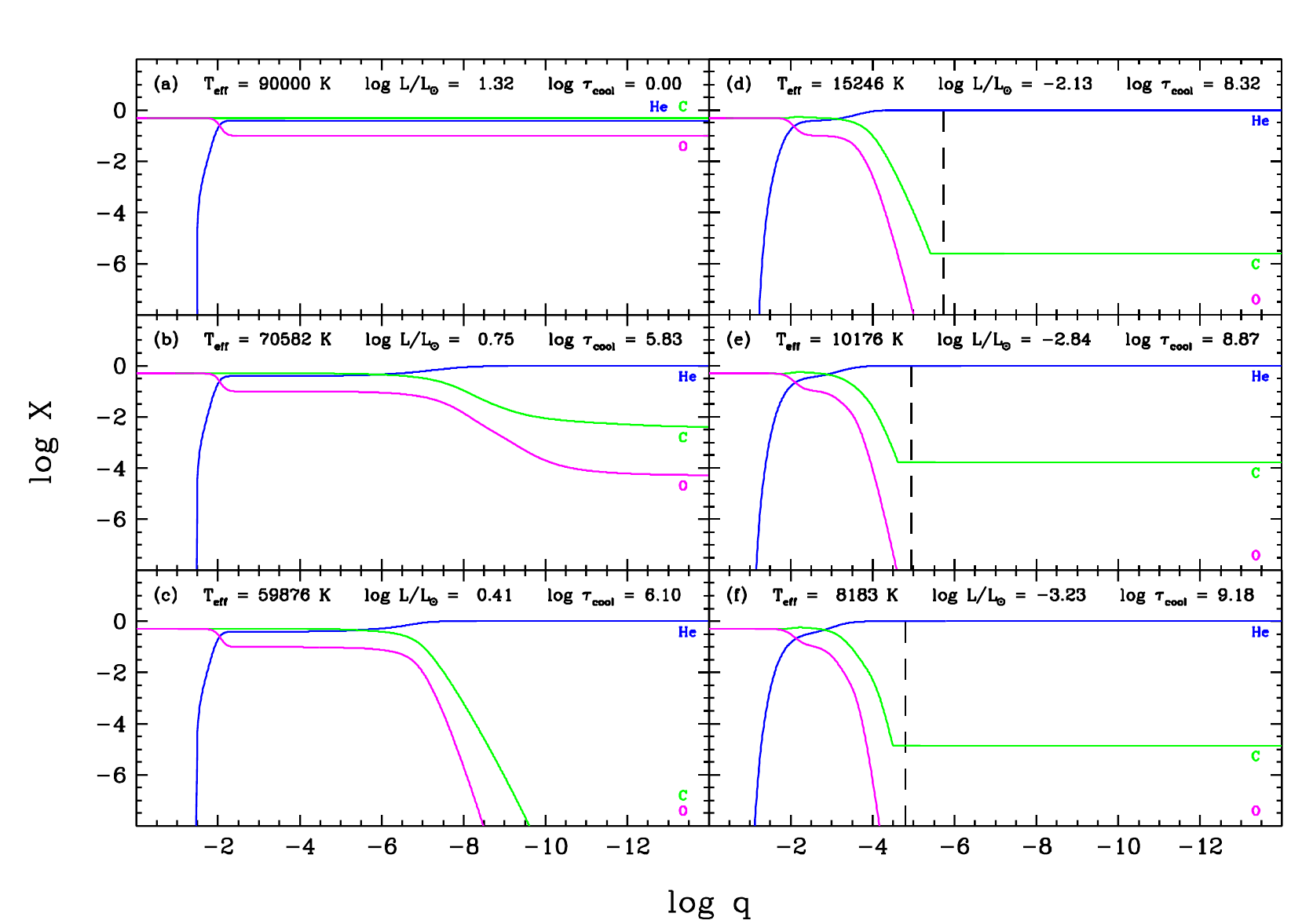}
\caption{Chemical structure at selected stages along our reference sequence. The elemental mass fraction abundances are shown as a function of the fractional mass depth ($q = 1 - m/M$). The hydrogen, helium, carbon, and oxygen abundance profiles are displayed as red, blue, green, and magenta curves, respectively. The location of the base of the convection zone is indicated by a dashed black line. The effective temperature, surface luminosity, and cooling age are given at the top of each panel.}
\vspace{2mm}
\label{fig:evol}
\end{figure*}

For completeness, we first briefly reproduce and discuss here the results of our standard simulation introduced in Section 3.1 of Paper II. Figure \ref{fig:evol} displays the run of the elemental mass fractions with depth at selected stages along the evolutionary sequence. The depth is measured as the fractional mass above the point of interest, $q = 1 - m/M$, where $m$ is the usual interior mass. This figure is essentially a reduced version of Figure 1 of Paper II (see also Appendix B of Paper II for an animation showing the entire simulation). Panels (a) to (c) illustrate the PG 1159-to-DO transition at high effective temperature: because the fading wind gradually loses its ability to compete efficiently against gravitational settling, the surface carbon and oxygen sink into deeper layers, and the initial PG 1159 star thereby becomes a pure-helium DO white dwarf. This transformation necessarily involves a short intermediate phase in which the model exhibits a helium-dominated yet mildly carbon-polluted atmosphere and would consequently appear as a DOZ white dwarf\footnote{In the canonical classification scheme, the letter Z indicates the presence of spectral features of any metal {\it except} carbon, which is identified distinctly by the letter Q \citep{wesemael1993}. Therefore, carbon-bearing DO white dwarfs, which exhibit weak C {\sc iv} lines in addition to the usual strong He {\sc ii} lines, should formally be assigned the spectral type DOQ. However, some carbon-polluted DO white dwarfs also show traces of other heavy elements, such as nitrogen and oxygen, and thus have historically been called DOZ stars \citep{dreizler1996}. In the present work, we continue to use the latter convention, that is, carbon-bearing DO white dwarfs are referred to as DOZ stars even if carbon is the only metal detected in their optical spectra.}. Besides, we note that the outer chemical profile of panel (c) is not entirely realistic: at this temperature, radiative levitation (which is not included in our calculations) is expected to support minute amounts of metals in the outer envelope \citep{chayer1995,dreizler1999}. During the DO and DB phases, the pure-helium layer grows as heavy elements diffuse further downward. Panels (d) to (f) outline the DB-to-DQ transition at low effective temperature: the superficial convection zone (the base of which is indicated by a dashed black line) expands and eventually comes into contact with the PG 1159-like layer at the bottom of the envelope, thereby causing carbon to be dredged up to the surface. The atmospheric carbon abundance of the resulting DQ white dwarf reaches a maximum and then decreases again due to an ionization effect below the convective region: the average ionization state of carbon changes, which causes the carbon diffusion tail to become steeper and thus to move away from the convection zone \citep{pelletier1986}. Note that the fully mixed region (which corresponds to the flat part of the carbon profile) extends beyond the formal convective boundary as a consequence of convective overshoot.

\begin{figure*}
\centering
\includegraphics[width=2.\columnwidth,clip=true,trim=2.0cm 9.25cm 1.25cm 10.5cm]{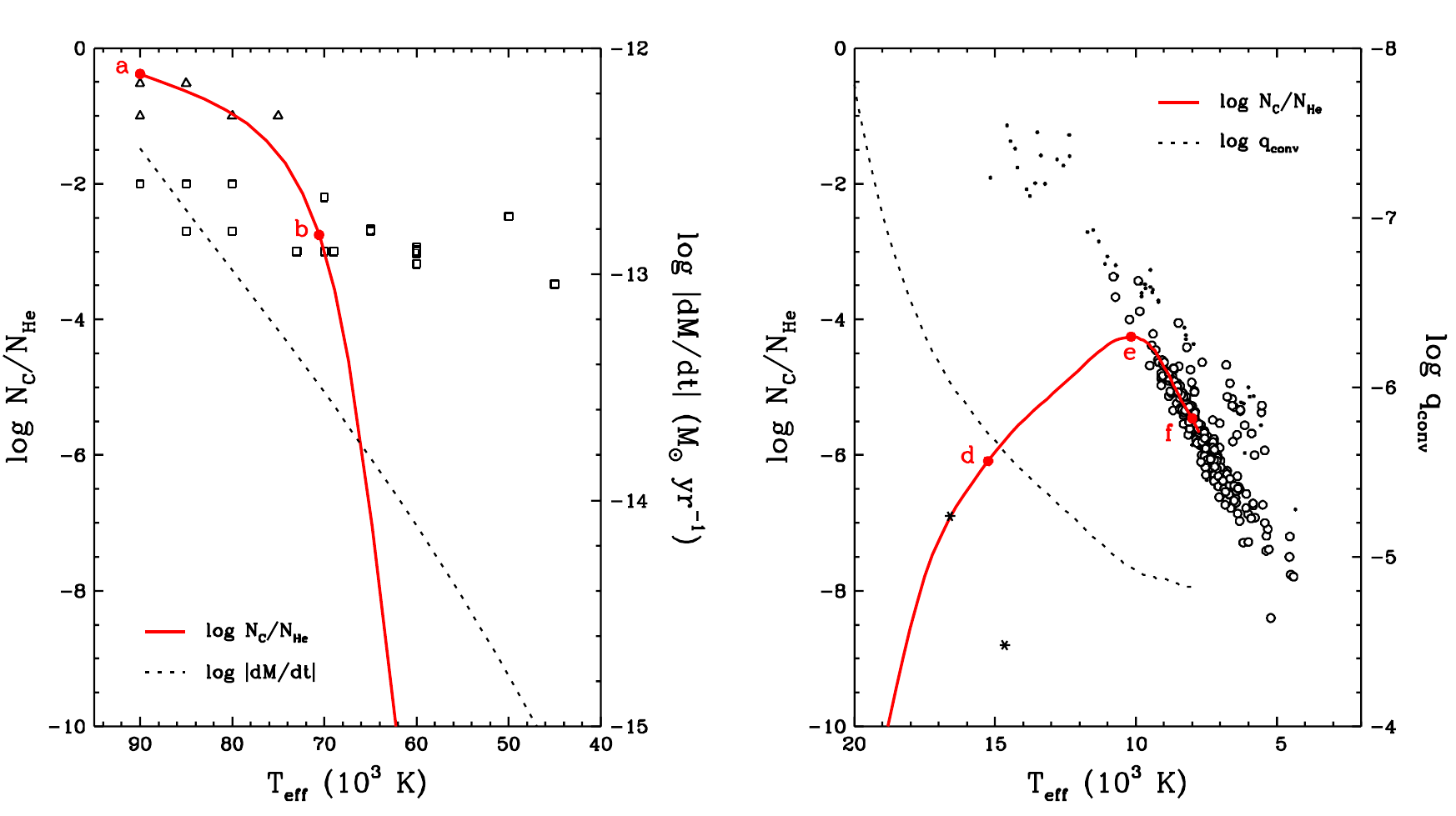}
\caption{Evolution of the atmospheric carbon-to-helium number ratio as a function of effective temperature in our reference sequence, shown as a red curve. The red points labeled by letters indicate the location of the models displayed in Figure \ref{fig:evol} along the curve. The dotted black line (read on the right axis) gives the wind mass-loss rate in the left panel and the fractional mass depth of the convection zone in the right panel. Also shown are empirical carbon abundance measurements taken from the literature for several types of objects: PG 1159 stars are represented as triangles \citep{werner2006,werner2014b}, DOZ white dwarfs as squares \citep{dreizler1996,werner2014a,reindl2014b}, DB white dwarfs as asterisks \citep{desharnais2008}, and DQ white dwarfs as large ($M \le 0.7 \ \msun$) or small ($M > 0.7 \ \msun$) circles \citep{coutu2019,blouin2019b}.}
\vspace{2mm}
\label{fig:full}
\end{figure*}

The chemical evolution of our reference model is summarized in Figure \ref{fig:full}, where the red line shows the surface carbon abundance (more precisely, the carbon-to-helium number ratio $N_{\rm C}/N_{\rm He}$) as a function of effective temperature. The red points labeled by letters along the curve correspond to the various stages displayed in Figure \ref{fig:evol}. The left and right panels cover the temperature ranges relevant to the PG 1159-to-DO and DB-to-DQ transitions, respectively. In each panel, the dotted black line (read on the right axis) shows a physical quantity related to the transport process responsible for the carbon abundance change. The left panel gives the mass-loss rate, which constitutes a measure of the capacity of the wind to prevent element sedimentation. We can see that the carbon abundance starts to drop abruptly once the mass-loss rate falls below a critical value of $\dot{M} \sim 10^{-13} \ \msun \ \rm{yr}^{-1}$, which occurs around $\Teff \sim 80,000$ K for the adopted mass-loss law. In our calculations, the carbon abundance subsequently falls to zero, but in reality, a small amount of carbon should be temporarily maintained at the surface by radiative levitation. The right panel displays the location of the base of the superficial convection zone in terms of fractional mass. The correlation between the carbon pollution and the inward expansion of the convective region is obvious.

Figure \ref{fig:full} also compares our predicted carbon abundance pattern with empirically measured carbon abundances for several types of white dwarfs. At high effective temperature, we show the coolest PG 1159 stars from \citet{werner2006} and \citet{werner2014b} as triangles, as well as the DOZ white dwarfs from \citet{dreizler1996}, \citet{werner2014a}, and \citet{reindl2014b} as squares. Note that these carbon-bearing DOZ stars represent only a subset of the whole DO population, which also comprises many pristine DO stars exhibiting pure helium-line spectra. In particular, the optical detection limit of carbon in a hot helium-rich atmosphere is $\log N_{\rm C}/N_{\rm He} \sim -3.0$ \citep{dreizler1996}, hence the near absence of objects below this value. Clearly, the population of DOZ white dwarfs cannot be fully explained by our standard sequence alone, a result that has been known for some time. For instance, as pointed out by \citet{reindl2014b}, the two coolest objects, SDSS J0301+0508 and SDSS J2239+2259, have surprisingly high carbon contents for their temperatures. One could allege that radiative levitation may be the solution to this puzzle, but the carbon abundances predicted by existing models of radiative levitation in hot white dwarfs are too low ($-4.5 \lta \log N_{\rm C}/N_{\rm He} \lta -3.5$; \citealt{chayer1995}). \citet{unglaub2000} argued that the observed scatter is simply a consequence of the fact that PG 1159 stars have different masses and compositions and therefore undergo the PG 1159-to-DO transition at different effective temperatures. Another possibility, alluded to by \citet{reindl2014b,reindl2014a}, is that DOZ white dwarfs actually descend from the so-called O(He) stars rather than from the PG 1159 stars. We discuss both ideas in light of our calculations in Section \ref{sec:disc_DO}. For now, we nevertheless note that the rudimentary wind model employed in our reference simulation successfully accounts for the disappearance of PG 1159 stars below $\Teff \sim 75,000$ K, as demonstrated before by \citet{unglaub2000} and \citet{quirion2012}. Moreover, it is also roughly consistent with our empirical knowledge of hot DAO white dwarfs, in which the wind is responsible for the helium pollution of the hydrogen-dominated atmosphere (\citealt{gianninas2010}; Paper I).

At low effective temperature, the normal-mass DQ white dwarfs ($M \le 0.7 \ \msun$) from \citet{coutu2019} and \citet{blouin2019b} are represented by large open circles; the remaining more massive objects are shown as small dots. Our model nicely reproduces the tight sequence formed by classical DQ stars, or at least the first half of it. At this time, we cannot push our calculations to lower effective temperature because of the unavailability of OPAL opacities for carbon in this physical regime. As mentioned in Paper II, this good agreement is not entirely coincidental: it was actually achieved by fine-tuning the overshoot parameter (assuming standard values for the other parameters), hence our choice of $f_{\rm ov} = 0.075$ as default value (see Section \ref{sec:res_DQ}). The absence of DQ white dwarfs on the ascending part of the theoretical curve is likely due to the optical visibility limit of carbon, which is $\log N_{\rm C}/N_{\rm He} \sim -4.5$ in this temperature range \citep{coutu2019}. In other words, the immediate precursors of classical DQ stars appear as genuine DB stars above $\Teff \sim 10,000$ K. However, their atmospheric carbon can still be detected in the ultraviolet. The two asterisks in Figure \ref{fig:full} represent the two cool DB white dwarfs ($\Teff < 20,000$ K) with ultraviolet carbon abundance measurements from \citet{desharnais2008}. One of them, GD 378, falls directly on the predicted curve and thereby seems entirely consistent with our current understanding of the DB-to-DQ transition, while the other one, GD 408, has an unexpectedly low carbon content. Nevertheless, these two objects also exhibit traces of other metals that probably originate from the accretion of planetary material \citep{desharnais2008,klein2020,klein2021}, thus it is not clear whether the detected carbon has been dredged up or accreted.

Finally, it is important to mention that although the PG 1159$-$DO$-$DB$-$DQ scenario satisfactorily describes the evolution of most carbon-polluted white dwarfs, it fails to explain the existence of two known classes of such objects. First, ultraviolet observations have revealed the presence of carbon at the surface of many hot DB stars ($\Teff > 20,000$ K; not shown here), similarly to their cooler counterparts but in even higher amounts ($\log N_{\rm C}/N_{\rm He} \sim -5.5$; \citealt{petitclerc2005,koester2014}). It is clear from Figure \ref{fig:full} that these objects cannot be accounted for by standard carbon dredge-up, and their origin remains uncertain to this day \citep{fontaine2005,koester2014}. Second, recent studies have confirmed the existence of a second DQ sequence in the $\log N_{\rm C}/N_{\rm He}-\Teff$ diagram populated mainly by the so-called hot and warm DQ white dwarfs, which are generally hotter, more massive, and more carbon-rich than the classical DQ white dwarfs \citep{coutu2019,koester2019}. These peculiar objects, which correspond to the small dots in Figure \ref{fig:full}, are strongly suspected to be merger products \citep{dunlap2015,coutu2019}, and thus the standard PG 1159$-$DO$-$DB$-$DQ evolutionary channel obviously does not apply in this case. It is beyond the scope of this work to investigate the chemical evolution of hot DB stars and of hot and warm DQ stars, and therefore these questions will be addressed elsewhere.

\subsection{The PG 1159-to-DO Transition} \label{sec:res_DO}

The PG 1159-to-DO transition is governed by the interplay between atomic diffusion and mass loss in the outermost layers of the envelope. Consequently, it depends mainly on the efficiency of gravitational settling and on the strength of the wind, which are respectively controlled by the total mass and the mass-loss rate. It is also reasonable to expect that the carbon abundance pattern is influenced by the initial carbon content. In contrast, other parameters such as the thickness of the envelope and the treatment of convective mixing play no role in the transformation. Besides, once the wind has faded, radiative levitation should sustain slight heavy-element pollution, but this phenomenon is disregarded in our models.

\begin{figure}
\centering
\includegraphics[width=\columnwidth,clip=true,trim=2.0cm 4.75cm 2.0cm 6.25cm]{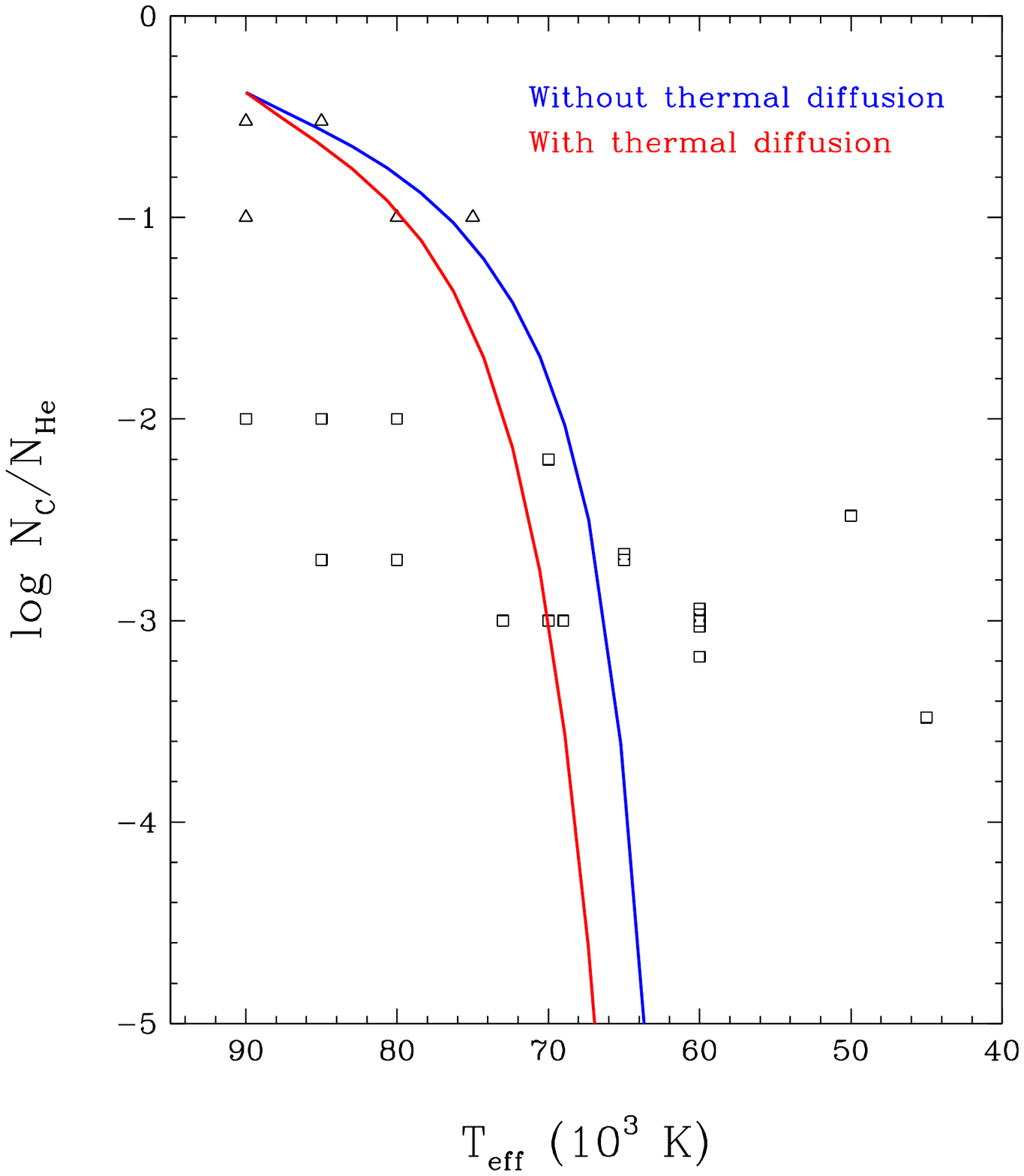}
\caption{Effect of thermal diffusion on the evolution of the atmospheric carbon-to-helium number ratio at high effective temperature. The red curve shows our reference sequence, which includes thermal diffusion, whereas the blue curve represents an analogous sequence ignoring thermal diffusion. Empirical carbon abundance measurements for PG 1159 stars \citep{werner2006,werner2014b} and DOZ white dwarfs \citep{dreizler1996,werner2014a,reindl2014b} are displayed as triangles and squares, respectively.}
\vspace{2mm}
\label{fig:DO_D}
\end{figure}

Before investigating these effects, we first briefly demonstrate the importance of thermal diffusion at the very high effective temperatures of PG 1159 and DO stars. This was already done by \citet{althaus2004}, who however focused mostly on the chemical structure of cooler DB white dwarfs ($\Teff \sim 20,000 - 30,000$ K). Figure \ref{fig:DO_D} compares our standard sequence with an analogous calculation in which thermal diffusion was turned off. Note that in this and all subsequent figures, the reference sequence is displayed as a red line. Because thermal diffusion acts in the same direction as gravitational settling, ignoring this process causes the carbon and oxygen to sink more slowly, with the consequence that the temperature of the PG 1159-to-DO transition is erroneously lowered by $\sim$3000 K. Thus, we reach the same conclusion as \citet{althaus2004}: thermal diffusion must be considered when modeling the spectral evolution of hot white dwarfs. This should be kept in mind when referring to previous calculations of the PG 1159-to-DO transition by \citet{unglaub2000} and \citet{quirion2012}, which only included gravitational settling.

\begin{figure}
\centering
\includegraphics[width=\columnwidth,clip=true,trim=2.0cm 4.75cm 2.0cm 6.25cm]{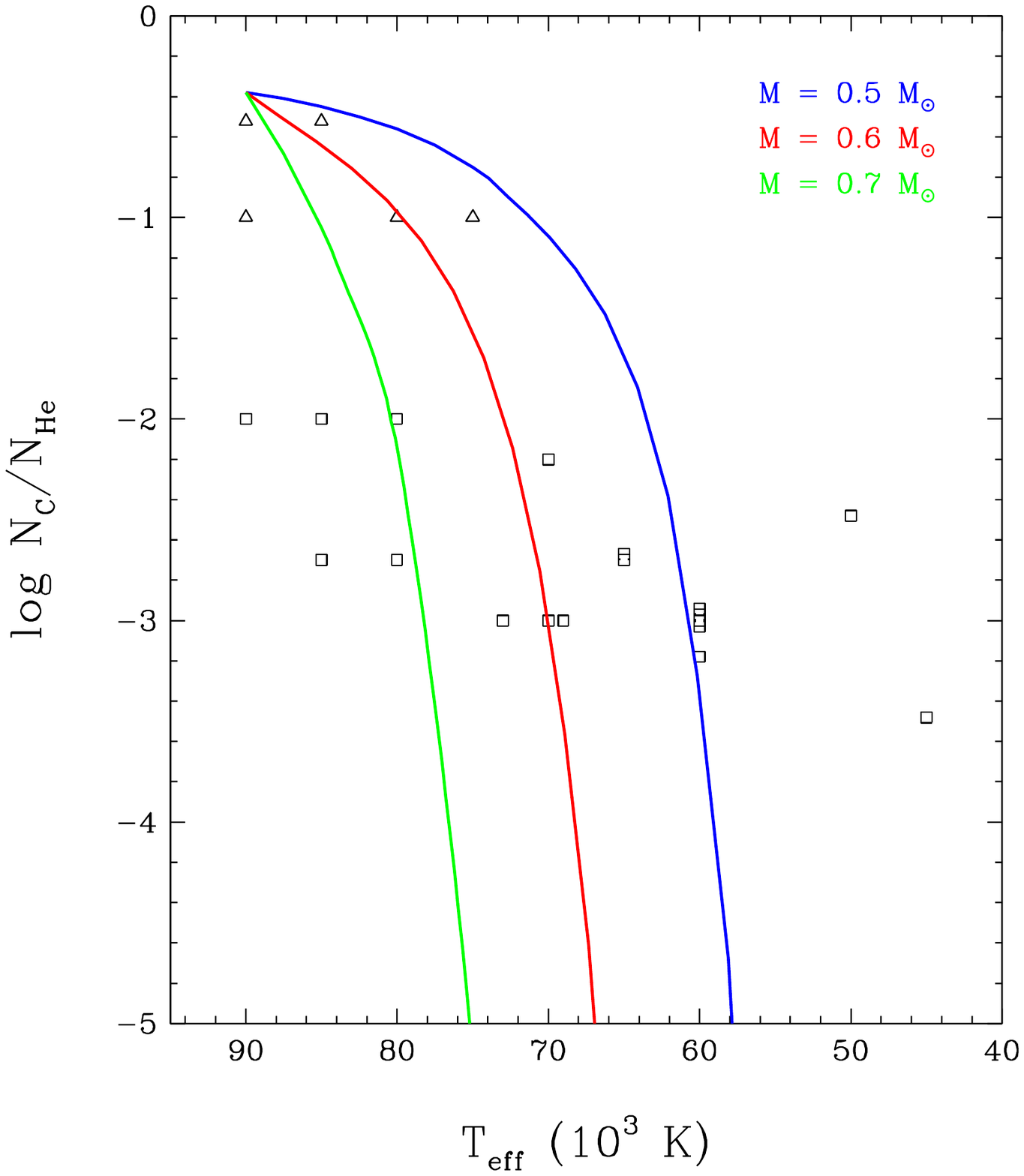}
\caption{Same as Figure \ref{fig:DO_D}, but for the effect of the stellar mass. The red curve shows our reference sequence, which assumes $M = 0.6 \ \msun$, whereas the blue and green curves represent analogous sequences assuming $M = 0.5$ and $0.7 \ \msun$, respectively.}
\vspace{2mm}
\label{fig:DO_M}
\end{figure}

Figure \ref{fig:DO_M} shows how varying the stellar mass impacts the evolution of the carbon abundance. Clearly, the more massive a PG 1159 star is, the earlier it transforms into a DO white dwarf. This behavior is partially due to the simple fact that a more massive star has a stronger gravitational field and thereby experiences more efficient diffusion. However, it must be realized that the mass also has an indirect influence on the strength of the wind: a more massive white dwarf is less luminous and therefore has a lower mass-loss rate by virtue of Equation \ref{eq:mdot}. Both effects act in the same direction: faster diffusion and weaker mass loss both accelerate the sedimentation of carbon and oxygen. The net result is significant: the effective temperature of the PG 1159-to-DO transition differs by $\sim$17,000 K between the models with $M = 0.5$ and $0.7 \ \msun$.  This relative difference is in good agreement with the works of \citet{unglaub2000} and \citet{quirion2012}, because the inclusion of thermal diffusion (which was ignored in these studies) affects all sequences in a similar way. Therefore, the large temperature dispersion that characterizes the observed carbon abundance pattern of DOZ stars may be explained in part by differences in mass from one object to another, which seems to support the existence of an evolutionary link between PG 1159 stars and DOZ white dwarfs. Nevertheless, one problem with this interpretation is that it predicts that the hottest DOZ white dwarfs should be the most massive, and vice versa, and yet such a temperature-mass relation is not observed. We come back to this point in Section \ref{sec:disc_DO}.

\begin{figure}
\centering
\includegraphics[width=\columnwidth,clip=true,trim=2.0cm 4.75cm 2.0cm 6.25cm]{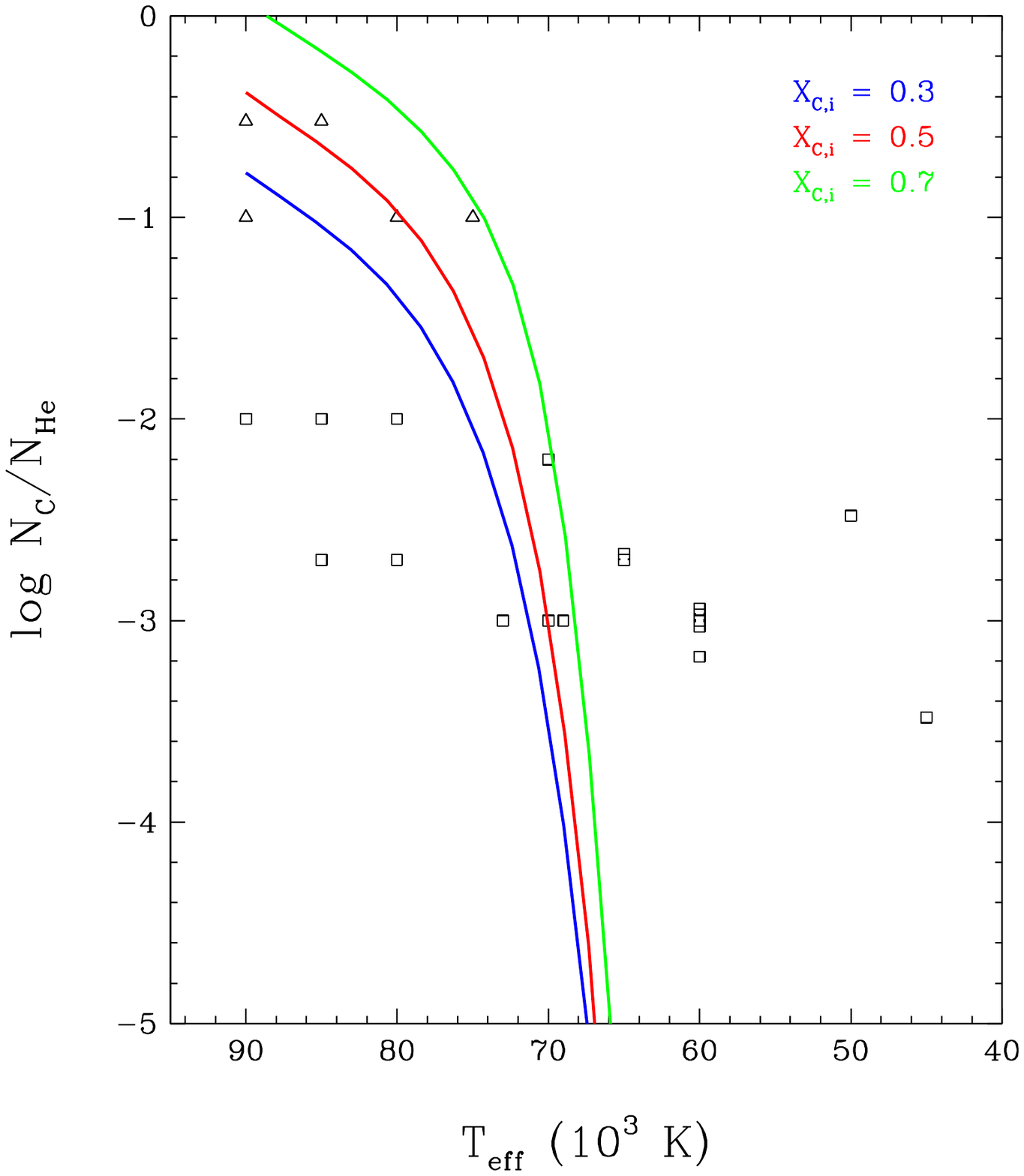}
\caption{Same as Figure \ref{fig:DO_D}, but for the effect of the initial carbon mass fraction in the envelope. The red curve shows our reference sequence, which assumes $X_{\rm C,i} = 0.5$, whereas the blue and green curves represent analogous sequences assuming $X_{\rm C,i} = 0.3$ and $0.7$, respectively.}
\vspace{2mm}
\label{fig:DO_XC}
\end{figure}

The influence of the initial carbon content on the PG 1159-to-DO transition is illustrated in Figure \ref{fig:DO_XC}. We compare our reference sequence with helium-rich, carbon-poor ($X_{\rm C,i} = 0.3$) and helium-poor, carbon-rich ($X_{\rm C,i} = 0.7$) PG 1159 models. While the carbon abundance is obviously altered at the onset of the evolution, the effect almost vanishes once the carbon starts sinking rapidly into the star. Hence, it appears that the diversity of surface composition seen among PG 1159 stars \citep{werner2006} does not translate directly into a scattered distribution of DOZ white dwarfs in the $\log N_{\rm C}/N_{\rm He}-\Teff$ diagram. That said, we recall that our simplistic wind model ignores the effect of the heavy element content on the mass-loss rate and thus on the ability of the wind to counteract gravitational settling. A higher carbon abundance is expected to produce a stronger wind and thereby a later PG 1159-to-DO transition \citep{unglaub2000}. For this reason, the gap between the sequences displayed in Figure \ref{fig:DO_XC} is likely underestimated.

\begin{figure}
\centering
\includegraphics[width=\columnwidth,clip=true,trim=2.0cm 4.75cm 2.0cm 6.25cm]{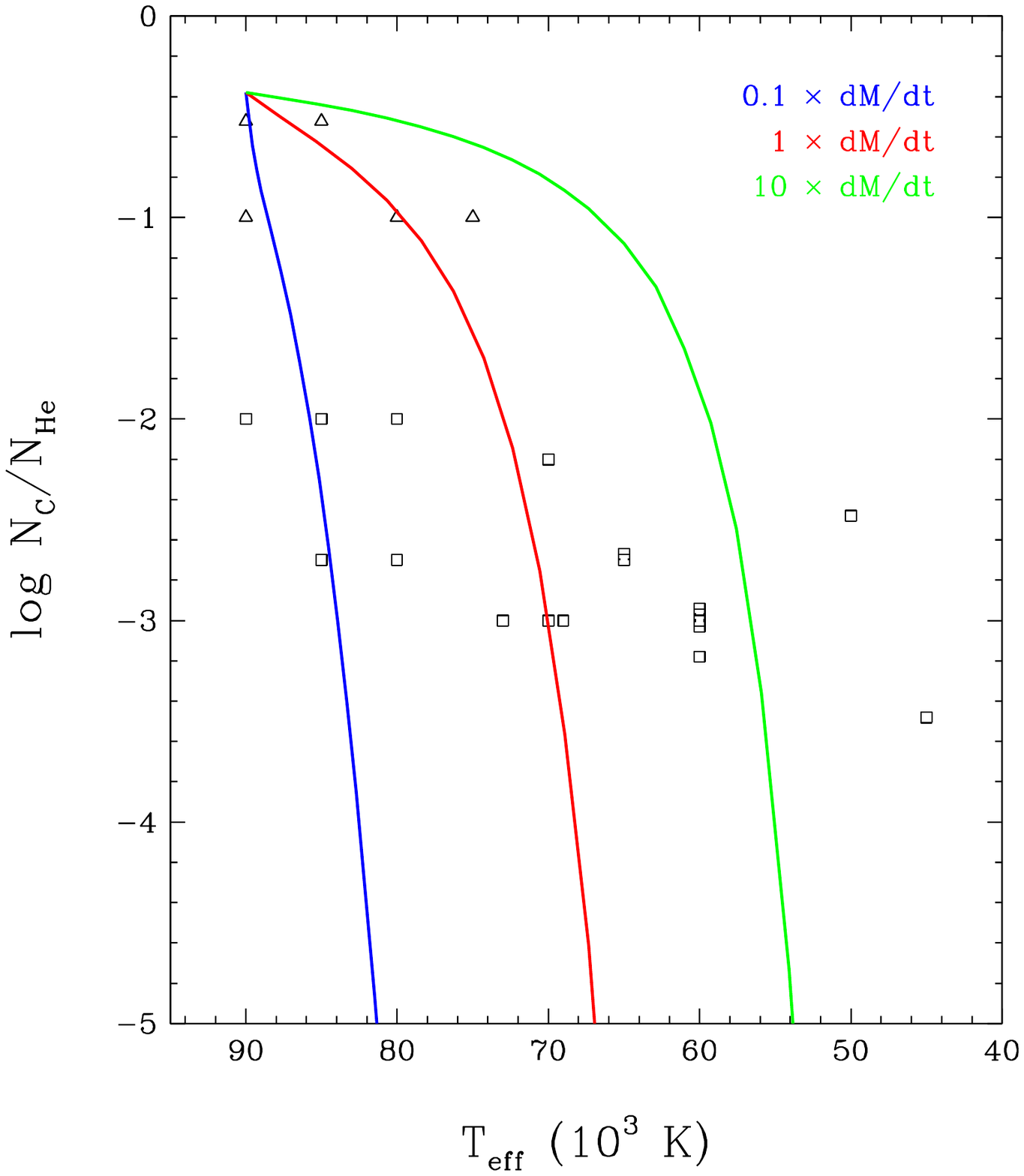}
\caption{Same as Figure \ref{fig:DO_D}, but for the effect of the mass-loss rate. The red curve shows our reference sequence, which uses Equation \ref{eq:mdot}, whereas the blue and green curves represent analogous sequences using Equation \ref{eq:mdot} divided and multiplied by a factor of 10, respectively.}
\vspace{2mm}
\label{fig:DO_W}
\end{figure}

Given this and other uncertainties regarding the treatment of the wind, we explore in Figure \ref{fig:DO_W} the impact of arbitrarily enhancing or reducing the mass-loss law by a factor of 10. For the low mass-loss rate, the wind is too weak to compete against downward diffusion, and thus the PG 1159 star turns into a DO white dwarf very quickly at the onset of the simulation. On the other hand, for the high mass-loss rate, the support mechanism remains highly efficient down to much lower effective temperature, such that the PG 1159-to-DO transition occurs around $\Teff \sim 60,000$ K. This behavior is similar to that reported by \citet{quirion2012}. Consequently, a wide range of mass-loss rates could in principle explain the existence of DOZ white dwarfs across a wide range of temperatures. However, there is no theoretical justification for such large variations (a factor of 100) in the strength of the wind from one star to another; in particular, the metallicity effect is expected to be much smaller \citep{unglaub2000}. Moreover, this unrealistic scenario still fails to account for the most extreme objects, such as SDSS J0301+0508 and SDSS J2239+2259 at low effective temperature. We further discuss the implications of our results for the origin of DOZ white dwarfs in Section \ref{sec:disc_DO}.

\subsection{The DB-to-DQ Transition} \label{sec:res_DQ}

The transformation of a DB white dwarf into a DQ white dwarf involves two competing transport mechanisms: atomic diffusion, which causes heavy elements to sink, and convective mixing, which brings some carbon back to the surface. Several stellar properties come into play in these phenomena. On one hand, the rate of carbon settling at the bottom of the envelope is controlled by the total mass and the envelope mass. On the other hand, the size of the fully mixed region is determined by the total mass, the mixing-length parameter, and the overshoot parameter. Furthermore, the magnitude of carbon dredge-up in a DQ star should depend to some extent on the carbon abundance of its PG 1159 progenitor. In the following, we investigate the quantitative effect of these parameters on the DB-to-DQ transition. Note that the wind present in the early evolution does not have any impact in later stages (it is completely ``forgotten'' by the star), hence it is irrelevant in this context. Recall that all figures display our standard simulation as a red line.

\begin{figure}
\centering
\includegraphics[width=\columnwidth,clip=true,trim=2.0cm 4.75cm 2.0cm 6.25cm]{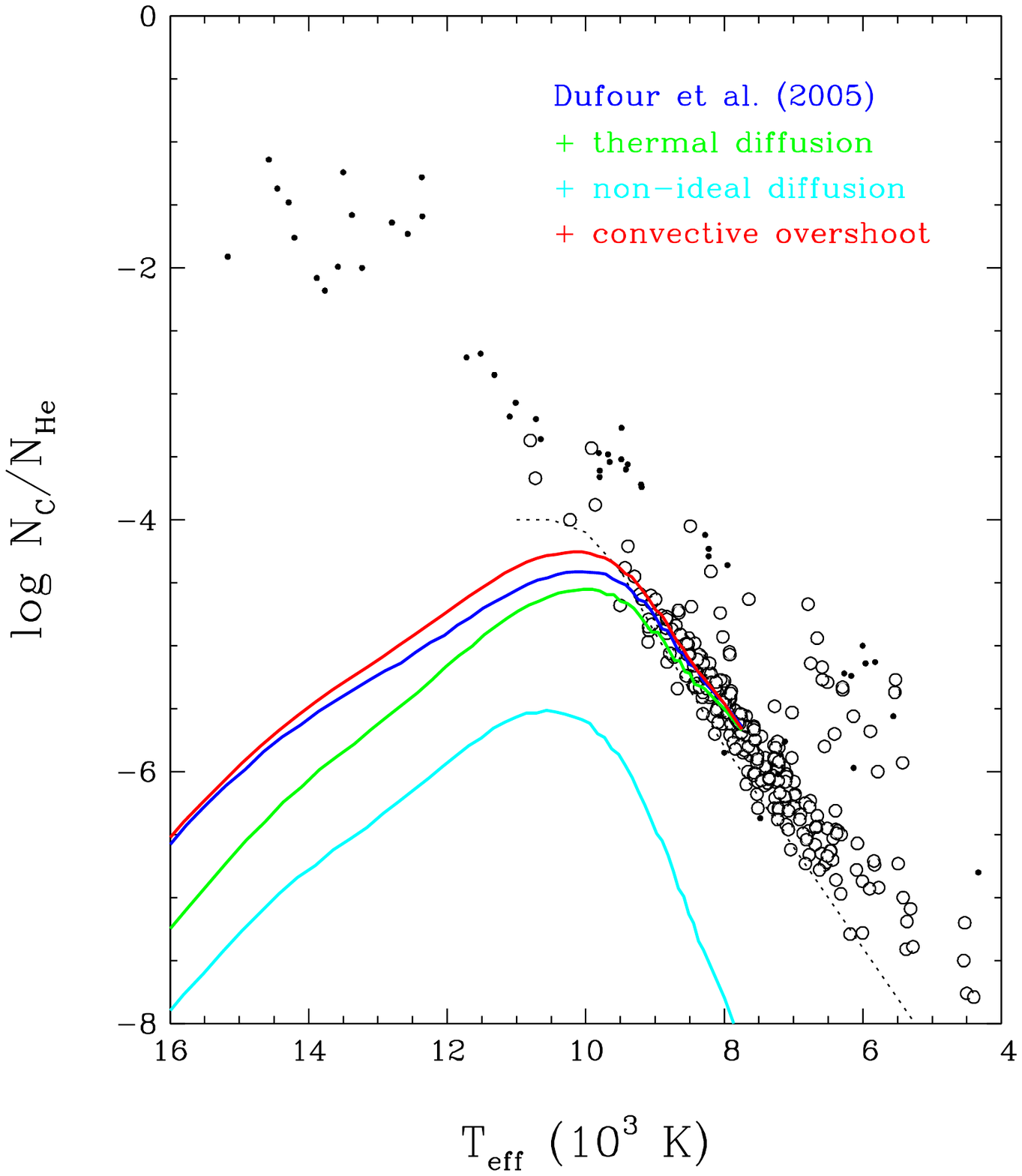}
\caption{Effect of various element transport processes on the evolution of the atmospheric carbon-to-helium number ratio at low effective temperature. The blue curve shows a sequence ignoring thermal diffusion, non-ideal diffusion, and convective overshoot, as in \citet{dufour2005}. The green, cyan, and red curves represent analogous sequences in which these processes are added one by one. The red line corresponds to our reference sequence, which includes all three effects. Empirical carbon abundance measurements for normal-mass ($M \le 0.7 \ \msun$) and high-mass ($M > 0.7 \ \msun$) DQ white dwarfs \citep{coutu2019,blouin2019b} are displayed as large circles and small dots, respectively. The dotted black line gives the optical detection limit of carbon for a signal-to-noise ratio ${\rm S/N} = 20$ based on the DQ model atmospheres of \citet{blouin2019a}.}
\vspace{2mm}
\label{fig:DQ_D}
\end{figure}

Before proceeding with our analysis, it is imperative to connect our current calculations with our previous set of DQ evolutionary models, first presented in Figure 12 of \citet{dufour2005} and more recently shown in Figure 12 of \citet{coutu2019}. The sequences published in \citet{dufour2005} neglect three element transport processes that are included in the present work: thermal diffusion, non-ideal diffusion, and convective overshoot. Figure \ref{fig:DQ_D} shows the evolution of the atmospheric carbon abundance before and during the DQ phase for various assumptions regarding these physical ingredients. Our starting point is the blue line, which represents a model computed within the theoretical framework adopted in \citet{dufour2005}. We see that this calculation nicely reproduces the empirical DQ sequence, as demonstrated in \citet{dufour2005} and \citet{coutu2019}\footnote{The blue curve in Figure \ref{fig:DQ_D} actually falls slightly higher in the $\log N_{\rm C}/N_{\rm He}-\Teff$ diagram compared to the analogous sequence of \citet{dufour2005}. The reason for this is that we use different values for the initial carbon mass fraction: \citet{dufour2005} assume $X_{\rm C,i} = 0.43$, whereas we adopt $X_{\rm C,i} = 0.5$, which results in a slightly higher final carbon abundance (see Figure \ref{fig:DQ_XC} and the associated discussion).}. The other curves illustrate the impact of successively adding thermal diffusion (green line), non-ideal diffusion (cyan line), and convective overshoot with $f_{\rm ov} = 0.075$ (red line, which corresponds to our reference simulation). Both thermal diffusion and non-ideal diffusion accelerate the sedimentation of carbon and thereby reduce the amount of carbon caught up by the convection zone and dredged up to the surface. According to Figure \ref{fig:DQ_D}, the two processes are of similar magnitude at $\Teff \sim 15,000$ K, but as the star cools, thermal diffusion becomes negligible while the effects of non-ideal diffusion grow stronger, as expected. Our calculations confirm that non-ideal diffusion is significant at the bottom of DQ white dwarf envelopes, as first pointed out by \citet{koester2020}. By further including moderate convective overshoot, which extends the homogeneously mixed region, the surface carbon abundance of the model can be brought back into agreement with the observations. In short, the good agreement obtained by \citet{dufour2005} is somewhat accidental: the effects of thermal and non-ideal diffusion are actually important but can be canceled out by convective overshoot.

\begin{figure}
\centering
\includegraphics[width=\columnwidth,clip=true,trim=2.0cm 4.75cm 2.0cm 6.25cm]{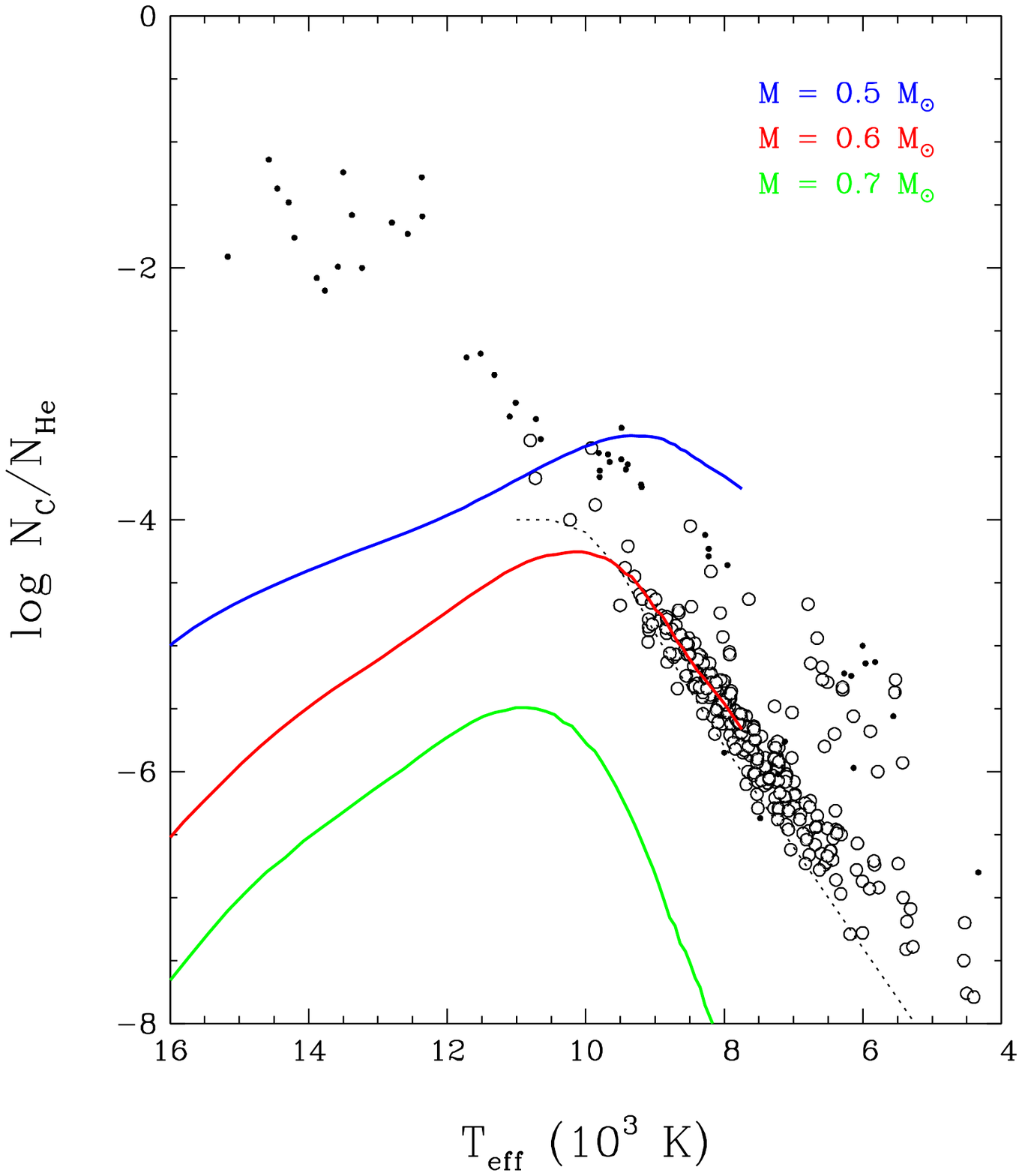}
\caption{Same as Figure \ref{fig:DQ_D}, but for the effect of the stellar mass. The red curve shows our reference sequence, which assumes $M = 0.6 \ \msun$, whereas the blue and green curves represent analogous sequences assuming $M = 0.5$ and $0.7 \ \msun$, respectively.}
\vspace{2mm}
\label{fig:DQ_M}
\end{figure}

Figure \ref{fig:DQ_M} illustrates the impact of varying the stellar mass while keeping all other parameters fixed to their default values. The carbon abundance pattern is highly sensitive to the stellar mass: at a given effective temperature, the carbon-to-helium number ratio differs by more than two orders of magnitude between the models with $M = 0.5$ and $0.7 \ \msun$. A more massive white dwarf undergoes less carbon pollution, primarily because its convection zone is shallower at a given temperature, and secondarily because its stronger gravitational field makes carbon settling more efficient. Note also that the various curves are shifted not only vertically, but also horizontally, such that the maximum carbon abundance is reached at slightly different temperatures. The mass dependence displayed in Figure \ref{fig:DQ_M} is very similar to that reported by \citet{pelletier1986}. In contrast, the calculations of \citet{camisassa2017} show a non-monotonic behavior with mass (see their Figure 10), but the fact that these models do not reproduce the decrease in carbon abundance at low effective temperature prevents us from making a more elaborate comparison. Taken at face value, our results seem to suggest that most DQ white dwarfs span an extremely narrow mass range; this idea is examined more closely in Section \ref{sec:disc_DQ_1}.

\begin{figure}
\centering
\includegraphics[width=\columnwidth,clip=true,trim=2.0cm 4.75cm 2.0cm 6.25cm]{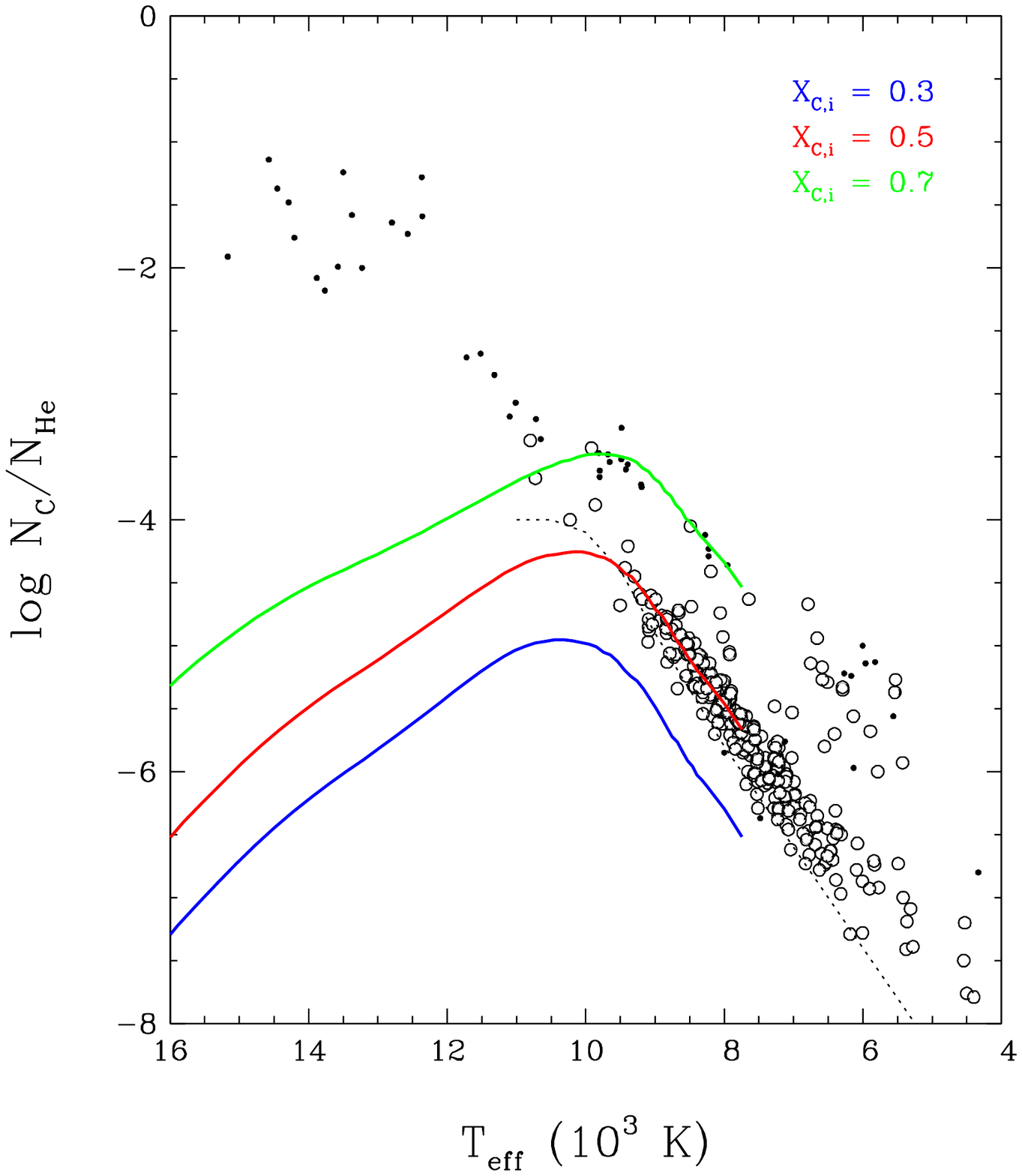}
\caption{Same as Figure \ref{fig:DQ_D}, but for the effect of the initial carbon mass fraction in the envelope. The red curve shows our reference sequence, which assumes $X_{\rm C,i} = 0.5$, whereas the blue and green curves represent analogous sequences assuming $X_{\rm C,i} = 0.3$ and $0.7$, respectively.}
\vspace{2mm}
\label{fig:DQ_XC}
\end{figure}

\begin{figure}
\centering
\includegraphics[width=\columnwidth,clip=true,trim=2.0cm 4.75cm 2.0cm 6.25cm]{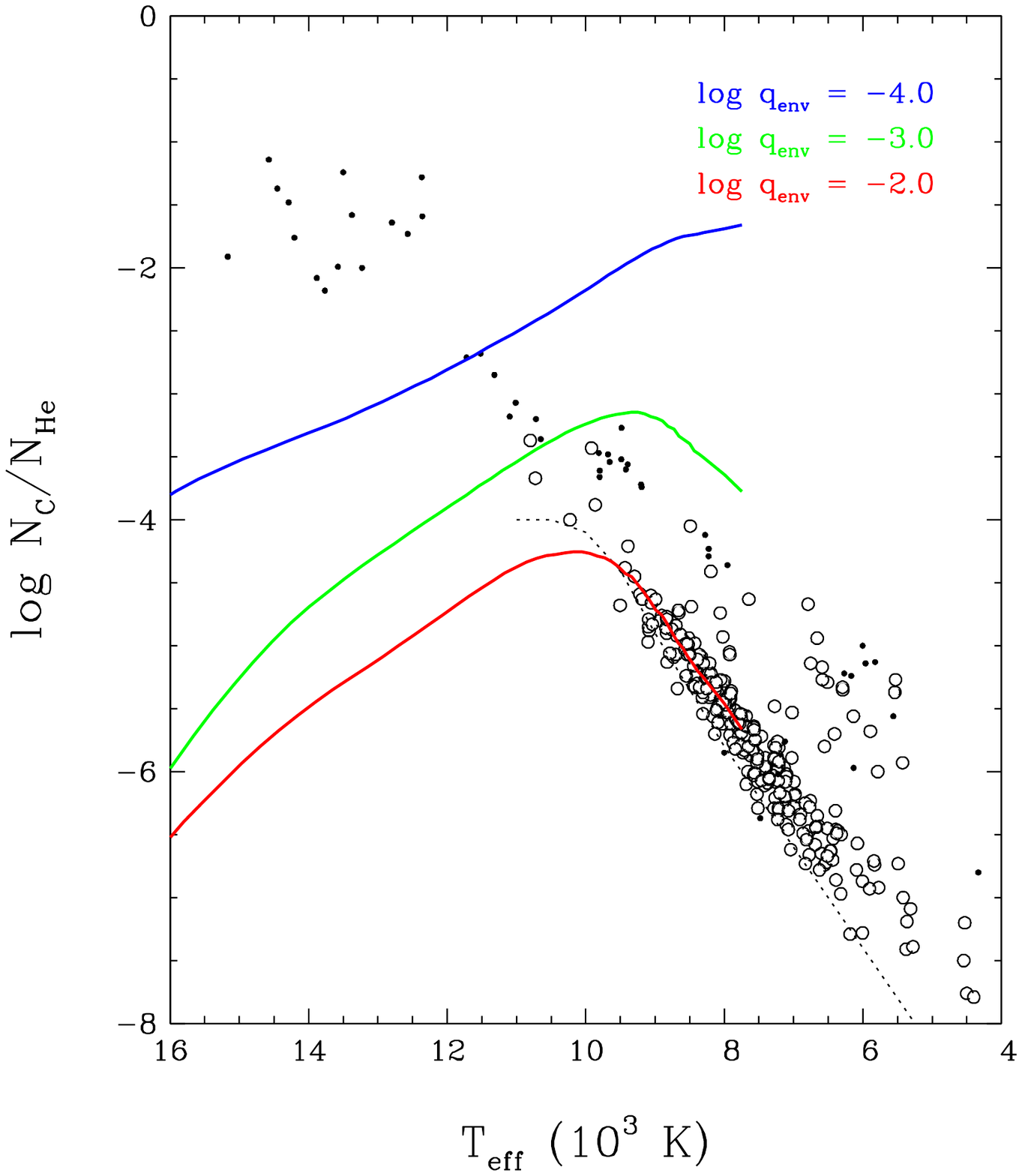}
\caption{Same as Figure \ref{fig:DQ_D}, but for the effect of the thickness of the envelope. The red curve shows our reference sequence, which assumes $\log q_{\rm env} = -2.0$, whereas the blue and green curves represent analogous sequences assuming $\log q_{\rm env} = -4.0$ and $-3.0$, respectively.}
\vspace{2mm}
\label{fig:DQ_LQ}
\end{figure}

Figures \ref{fig:DQ_XC} and \ref{fig:DQ_LQ} show how the chemical evolution is affected by the main envelope parameters, namely, the initial mass fraction of carbon and the thickness of the envelope. The interpretation of Figure \ref{fig:DQ_XC} is quite straightforward: the higher the carbon content of the PG 1159 progenitor, the higher the atmospheric carbon abundance of the DQ progeny. Figure \ref{fig:DQ_LQ} shows that thinner envelopes give rise to markedly enhanced carbon enrichment, a trend which is well known since the work of \citet{pelletier1986}. On the other hand, the results for $\log q_{\rm env} > -2.0$ (not displayed here) are strictly identical to those for $\log q_{\rm env} = -2.0$. This dual behavior can be understood as follows. For $\log q_{\rm env} \lta -2.5$, the diffusive timescales at the base of the envelope are short enough that element separation is fully achieved before the onset of convective dredge-up. In other words, the PG 1159-like mantle at the bottom of the envelope is completely disintegrated by diffusion, leaving only a pure-helium layer atop a carbon/oxygen core. In this situation, it is from the core that carbon is brought up to the surface, and thus the magnitude of carbon pollution depends sensitively on the location of the core/envelope boundary. In contrast, for $\log q_{\rm env} \gta -2.5$, the PG 1159-like layer remains intact throughout the DQ phase, and therefore the surface carbon abundance becomes insensitive to the thickness of the envelope. Like the mass in Figure \ref{fig:DQ_M}, we note that Figures \ref{fig:DQ_XC} and \ref{fig:DQ_LQ} seem to indicate that the DQ population is characterized by an extremely small range of $X_{\rm C,i}$ and $q_{\rm env}$ values, in apparent contradiction with the diversity of carbon abundances measured in PG 1159 stars \citep{werner2006}. We come back to this point in Section \ref{sec:disc_DQ_1}.

\begin{figure}
\centering
\includegraphics[width=\columnwidth,clip=true,trim=2.0cm 4.75cm 2.0cm 6.25cm]{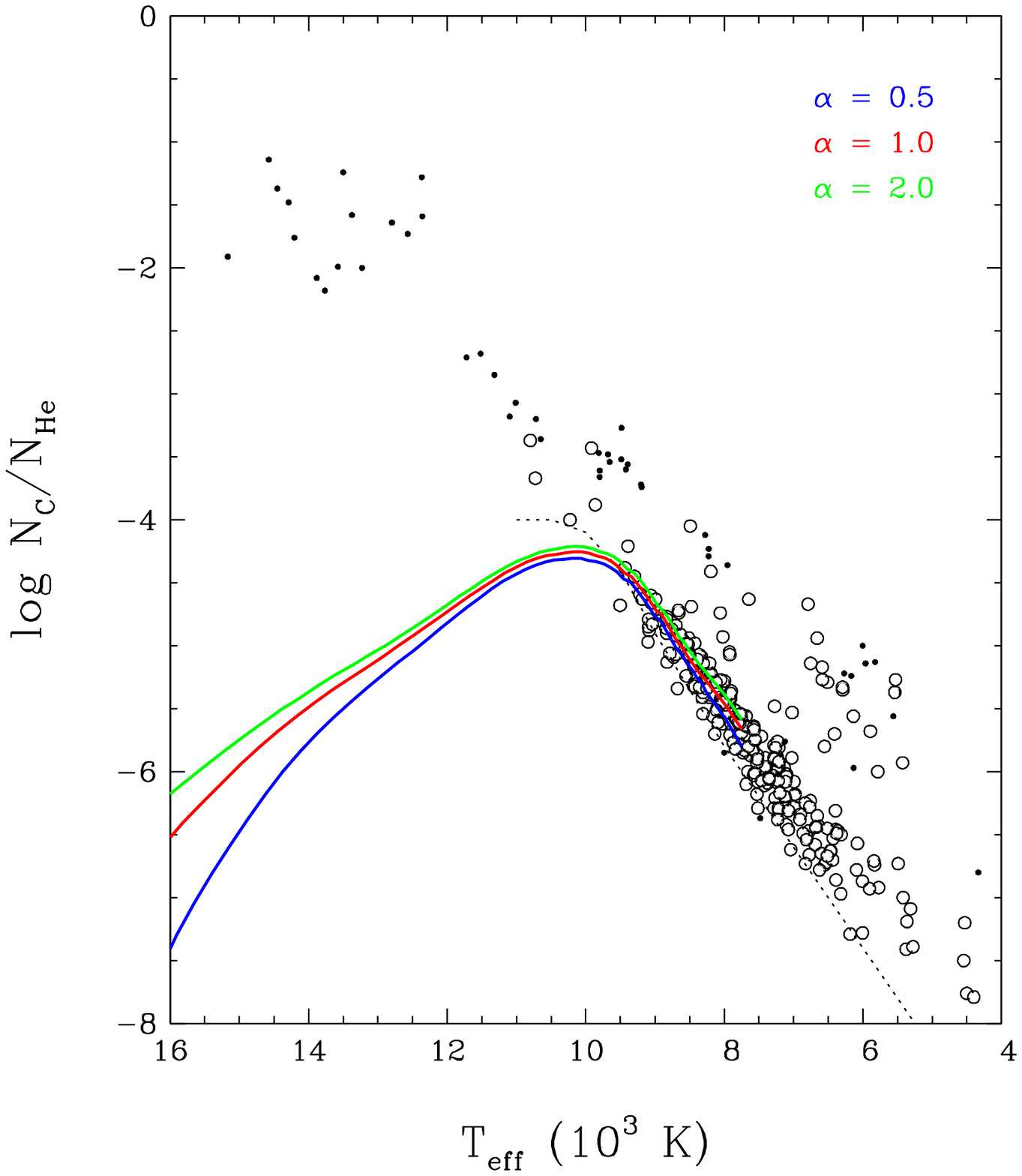}
\caption{Same as Figure \ref{fig:DQ_D}, but for the effect of the mixing-length parameter. The red curve shows our reference sequence, which assumes $\alpha = 1.0$, whereas the blue and green curves represent analogous sequences assuming $\alpha = 0.5$ and $2.0$, respectively.}
\vspace{2mm}
\label{fig:DQ_CV}
\end{figure}

\begin{figure}
\centering
\includegraphics[width=\columnwidth,clip=true,trim=2.0cm 4.75cm 2.0cm 6.25cm]{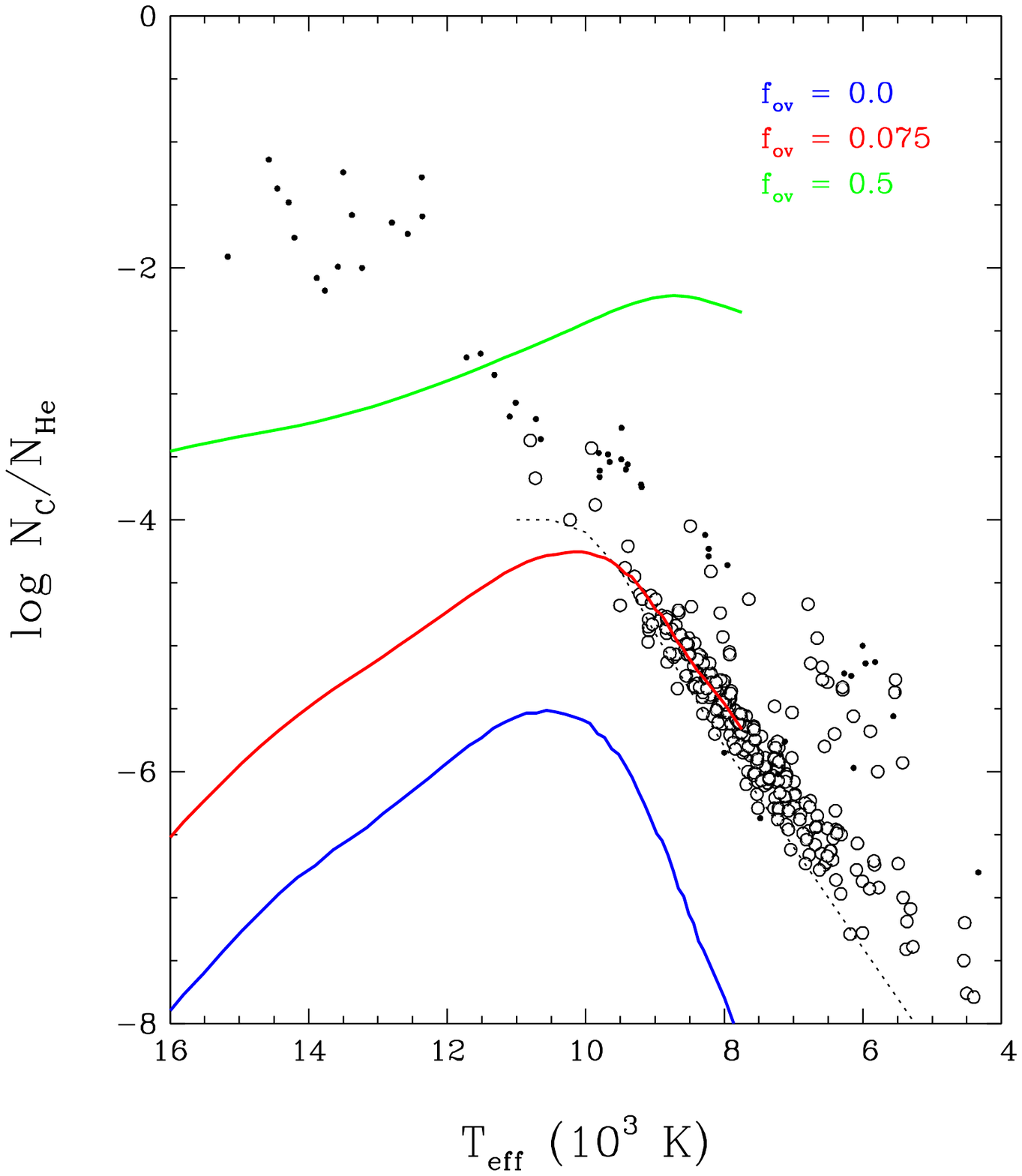}
\caption{Same as Figure \ref{fig:DQ_D}, but for the effect of the overshoot parameter. The red curve shows our reference sequence, which assumes $f_{\rm ov} = 0.075$, whereas the blue and green curves represent analogous sequences assuming $f_{\rm ov} = 0.0$ and $0.5$, respectively.}
\vspace{2mm}
\label{fig:DQ_OV}
\end{figure}

Figures \ref{fig:DQ_CV} and \ref{fig:DQ_OV} illustrate how the atmospheric composition is influenced by the treatment of convective mixing, which involves the mixing-length and overshoot parameters. Figure \ref{fig:DQ_CV} simply reaffirms the conclusion of \citet{pelletier1986} that the adopted version of the mixing-length theory is inconsequential at low effective temperature, because convection becomes essentially adiabatic. Nevertheless, the same cannot be said of the extent of convective overshoot. This physical feature is definitely the most poorly understood of all those studied so far, and its effect on the chemical evolution of DQ white dwarfs has seldom been investigated before. Figure \ref{fig:DQ_OV} demonstrates that the impact of the overshoot parameter is substantial. The model without overshoot ($f_{\rm ov} = 0.0$) exhibits a carbon abundance more than one order of magnitude lower than the reference model at a given temperature (this was already apparent in Figure \ref{fig:DQ_D}). The other theoretical curve displayed in Figure \ref{fig:DQ_OV} assumes $f_{\rm ov} = 0.5$, which is roughly the largest possible value according to recent hydrodynamical simulations of convection in warm DA white dwarfs \citep{kupka2018,cunningham2019}. In this sequence, the extensive overshoot leads to a very strong carbon enrichment, in fact much stronger than observed in cool DQ stars. All in all, the models with $f_{\rm ov} = 0.0$ and $0.5$ differ by three to five orders of magnitude in surface carbon abundance depending on the effective temperature.

It thus appears that the extent of convective overshoot is a sizable source of uncertainty in our simulations of the chemical evolution of DQ white dwarfs. However, we can turn the problem around: given standard values of the other important parameters, the overshoot parameter can be calibrated using the observed DQ sequence. This is precisely how we chose the $f_{\rm ov}$ value of our reference simulation: assuming $M = 0.6 \ \msun$, $\log q_{\rm env} = -2.0$, and $X_{\rm C,i} = 0.5$, we find that $f_{\rm ov} = 0.075$ yields the best match to the empirical $\log N_{\rm C}/N_{\rm He}-\Teff$ relation. We want to stress that this is not a firm determination, since our assumptions on $M$, $q_{\rm env}$, and $X_{\rm C,i}$ may not be perfectly accurate. Nevertheless, our approach can still provide some insight into the significance of convective overshoot in cool white dwarfs. This topic is further discussed in Section \ref{sec:disc_DQ_1}.

\subsection{The Effect of Residual Hydrogen} \label{sec:res_H}

In this section, we present one last simulation which differs from our standard sequence only by the addition of trace hydrogen in the envelope of the initial model ($X_{\rm H,i} = 10^{-4}$ down to $\log q = -4.0$, for a total hydrogen mass of $10^{-8} M$). We describe this calculation separately and in more detail because the presence of residual hydrogen significantly alters the chemical evolution. This simulation is similar to that presented in Section 3.2 of Paper II, except that the latter was started from an almost pure-helium DO white dwarf, whereas we consider here a PG 1159 star with large amounts of carbon and oxygen. This allows us to study for the first time the effect of hydrogen on the phenomenon of carbon dredge-up at low effective temperature.

\begin{figure*}
\centering
\includegraphics[width=2.\columnwidth,clip=true,trim=2.0cm 6.5cm 2.0cm 8.0cm]{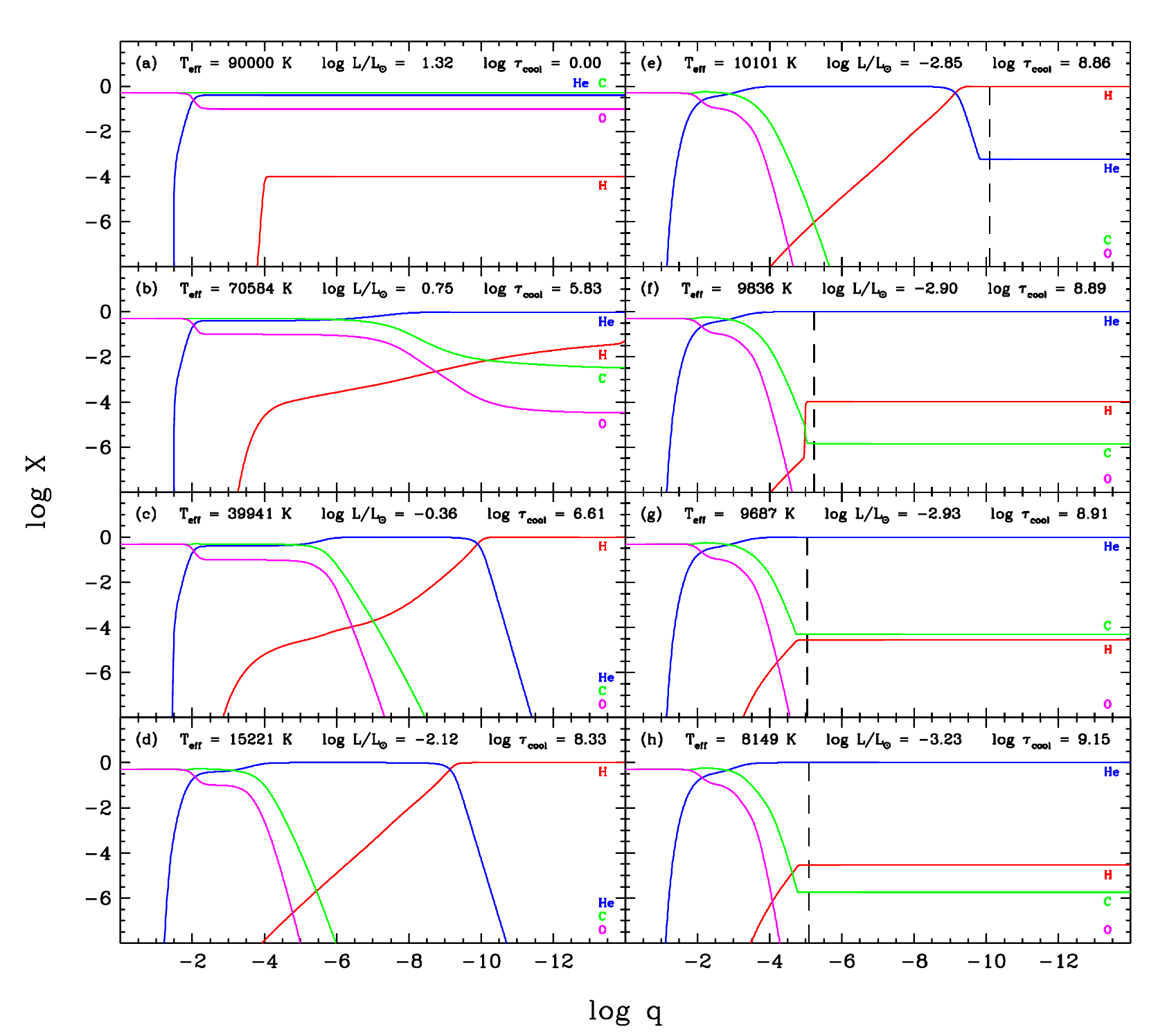}
\caption{Same as Figure \ref{fig:evol}, but for a sequence including residual hydrogen with an initial mass fraction $X_{\rm H,i} = 10^{-4}$.}
\vspace{2mm}
\label{fig:evol_H}
\end{figure*}

Figure \ref{fig:evol_H} shows snapshots of the chemical profile along our evolutionary sequence including hydrogen, similarly to Figure \ref{fig:evol} for our reference sequence. As can be seen in panels (a) to (d), the sedimentation of carbon and oxygen at high effective temperature is essentially unaffected by the presence of trace hydrogen. However, a crucial difference is that the hydrogen gradually diffuses upward, eventually forming a thin pure-hydrogen layer at the surface. Consequently, the PG 1159 star transforms briefly into a DO white dwarf and then into a DA white dwarf. For the parameters adopted here, the DO-to-DA transition occurs at $\Teff \sim 50,000$ K (see Section 3.2 of Paper II for details). It can be seen in panels (c) and (d) that the DA star harbors a triple-layered envelope, with hydrogen-rich, helium-rich and PG 1159-like layers from top to bottom, which is qualitatively consistent with previous calculations by \citet{althaus2005b,althaus2020a}. At $\Teff \sim 15,000$ K, all of the residual hydrogen has reached the surface, and thus the hydrogen and helium layers are in diffusive equilibrium. Besides, note that at this point, the total hydrogen mass has decreased to $\sim$10$^{-9} M$, because much of the outermost hydrogen-rich material has been ejected by the stellar wind. Furthermore, it is possible that the inclusion of nuclear burning in our calculations would have reduced the hydrogen content even further, as chemical diffusion initially carries some of the hydrogen to great depths (down to $\log q \sim -3.0$) where it should burn \citep{althaus2005b,renedo2010}.

The presence of a superficial pure-hydrogen layer has major repercussions on the subsequent chemical evolution. Going back to Figure \ref{fig:evol}, we notice that at $\Teff \sim 15,000$ K, our standard model has already begun to undergo carbon dredge-up, by virtue of the deep helium convection zone. In contrast, at the same effective temperature, the outer parts of our model including hydrogen remain devoid of carbon, since the thin hydrogen envelope is still convectively stable. Therefore, if a PG 1159 star contains enough residual hydrogen to become a DA rather than DB white dwarf, the onset of carbon dredge-up is delayed. However, the hydrogen layer does eventually become convective. As the star cools, the convection zone expands inward and reaches into the underlying helium reservoir at $\Teff \sim 10,000$ K, thereby mixing some helium into the hydrogen layer, as illustrated in panel (e) of Figure \ref{fig:evol_H}. This leads to a runaway mixing event: the higher the surface helium abundance, the deeper the convective region, the more helium is brought to the surface, and so on. The hydrogen very rapidly ends up being completely diluted into a helium-dominated envelope (see Section 3.2 of Paper II for details)\footnote{Note that the effective temperature of the convective mixing event is $\sim$200 K lower than in the simulation of Section 3.2 of Paper II. The reason for this is that the initial envelope composition adopted here is slightly different (we include carbon and oxygen), which results in a little less hydrogen being ejected by the wind at high temperature, and thus the remaining hydrogen layer at low temperature is very slightly thicker.}. The depth of the convection zone is now such that carbon from the PG 1159-like layer is dredged up to the surface. Because the hydrogen/helium mixing event and the ensuing growth of the convection zone occur on an extremely short timescale, the mass fraction of carbon in the atmosphere goes from 0 to $\sim$10$^{-4}$ almost instantaneously near $\Teff \sim 10,000$ K, as can be seen in panels (f) and (g). Finally, panel (h) shows that the carbon abundance then decreases with cooling, as in the hydrogen-free case. Interestingly, we note that the final hydrogen and carbon abundances are both just below their optical detection limit, and thus the model displayed in panel (h) would not be classified as a DAQ, DQA, DA, or DQ white dwarf, but rather as a plain DC white dwarf.

\begin{figure}
\centering
\includegraphics[width=\columnwidth,clip=true,trim=2.0cm 4.75cm 2.0cm 6.25cm]{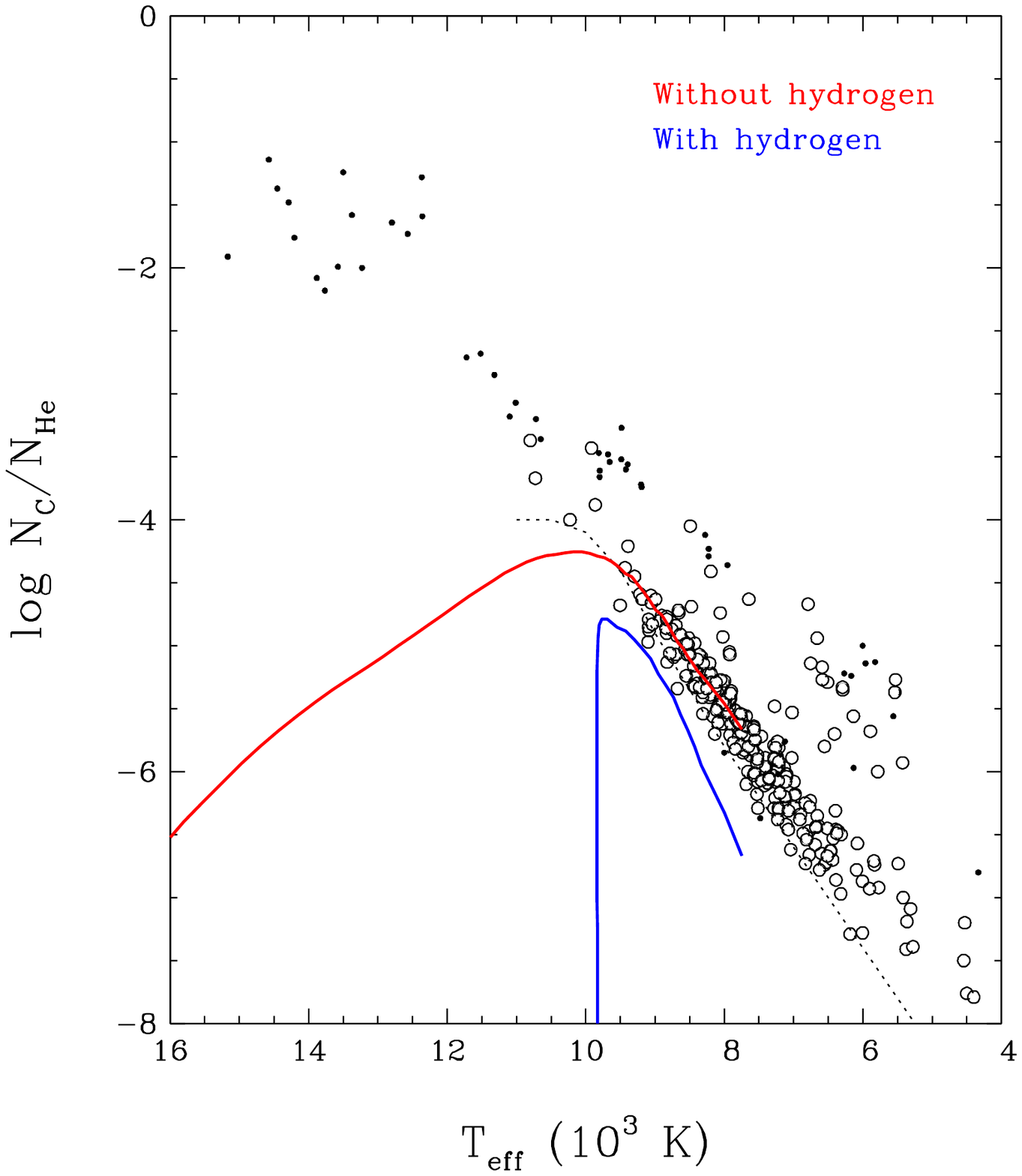}
\caption{Same as Figure \ref{fig:DQ_D}, but for the effect of residual hydrogen. The red curve shows our reference sequence, which does not include any hydrogen, whereas the blue curve represents an analogous sequence including residual hydrogen with an initial mass fraction $X_{\rm H,i} = 10^{-4}$.}
\vspace{2mm}
\label{fig:DQ_H}
\end{figure}

Figure \ref{fig:DQ_H} compares our sequences with and without hydrogen in the $\log N_{\rm C}/N_{\rm He}-\Teff$ diagram. The presence of residual hydrogen radically modifies the carbon abundance pattern: the increase in carbon abundance is very abrupt rather than gradual. In our calculation, this sudden carbon enrichment occurs at $\Teff \sim 10,000$ K, but this value is expected to depend on the amount of hydrogen: the thicker the superficial hydrogen layer, the lower the hydrogen/helium mixing temperature, the lower the carbon dredge-up temperature \citep{rolland2018,cunningham2019}. Furthermore, on the descending part of the theoretical curve, the carbon abundance is somewhat lower when hydrogen is included. This is because the presence of trace hydrogen in the envelope slightly reduces the extent of the convection zone, as is apparent when comparing the last panels of Figures \ref{fig:evol} and \ref{fig:evol_H}. Nevertheless, a very small adjustment of the overshoot parameter would be sufficient to bring our sequence with hydrogen in agreement with the observations. In this context, one could be tempted to deduce that perhaps most DQ white dwarfs descend from DA stars with thin hydrogen layers rather than from DB stars. However, we believe that this is highly unlikely. The reason is that such a DA-to-DQ transition involves a $\sim$0.5-Gyr intermediate phase in which the star appears as a DQA white dwarf, and it is a well-established observational fact that cool DQA white dwarfs are exceedingly rare \citep{dufour2011,coutu2019}. Therefore, we think that this particular spectral evolution channel only applies to a minority of DQ stars.

\section{Discussion} \label{sec:disc}

\subsection{The Origin of DOZ White Dwarfs} \label{sec:disc_DO}

The transformation of PG 1159 stars into DO white dwarfs is inevitable, given that carbon and oxygen must ultimately sink out of sight under the influence of gravitational settling. However, this does not mean that all DO white dwarfs necessarily descend from PG 1159 stars. In particular, the origin of the subset of DO white dwarfs exhibiting traces of carbon is a question that remains unsettled. Are they simply transitional objects currently in between the PG 1159 and DO phases, or are they the products of an entirely different evolutionary channel?

Based on their element transport calculations, \citet{unglaub2000} argued that PG 1159 stars with various masses and compositions transform into DO white dwarfs at various effective temperatures, thereby naturally explaining the large dispersion of the carbon abundance pattern of DOZ white dwarfs. We can take a fresh look at this idea thanks to our improved chemical evolution simulations and to the increased number of empirical carbon abundance determinations available in the literature. We confirmed the result of \citet{unglaub2000} that a relatively small change in the assumed stellar mass leads to a relatively large change in the PG 1159-to-DO transition temperature (that is, a difference of $\sim$17,000 K between $M = 0.5$ and $0.7 \ \msun$). We showed that part (but not all) of the observational scatter in the $\log N_{\rm C}/N_{\rm He}-\Teff$ diagram might be due to the intrinsic width of the DO mass distribution. Furthermore, we found that variations in the adopted wind mass-loss rate (which are theoretically expected from variations in the heavy element content) also have a considerable impact on the PG 1159-to-DO transition temperature. Nevertheless, for a given mass, the empirical data can be explained only if individual objects have mass-loss rates differing by more than a factor of 100, which is unreasonably high. Still, Figures \ref{fig:DO_M} and \ref{fig:DO_W} suggest that most of the measured carbon abundances can be accounted for with realistic parameters if we consider both the mass and wind effects simultaneously. This is reasonable since the PG 1159 population is characterized by both a range of masses and a range of metallicities at a given mass \citep{werner2006}. Therefore, according to our calculations, it remains plausible that most DOZ white dwarfs originate from PG 1159 stars.

However, there are still of few issues with this scenario. For instance, if the carbon abundance pattern does reflect the width of the DO mass distribution, then Figure \ref{fig:DO_M} indicates that the hottest DOZ white dwarfs should be the most massive, and vice versa. Such a temperature-mass relation is not observed in spectroscopic studies \citep{dreizler1996,werner2014a,reindl2014b}, which casts doubts on the above interpretation. Moreover, the coolest PG 1159 and DOZ stars are separated by a surprisingly large temperature gap. On one hand, the coolest known PG 1159 star, HS 0704+6153, has $\Teff = 75,000$ K \citep{werner2006}, and our models predict that it should turn into a pure-helium DO white dwarf before it reaches $\Teff = 60,000$ K (see Figure \ref{fig:DO_M}). On the other hand, DOZ white dwarfs exist down to much lower effective temperatures: SDSS J0301+0508 and SDSS J2239+2259 have $\Teff = 50,000$ and $45,000$ K, respectively \citep{reindl2014b}. Our calculations indicate that a very high mass-loss rate is likely required to produce such objects. Yet a very high mass-loss rate should also produce PG 1159 stars cooler than $\Teff \sim 75,000$ K, in contradiction with the observations (see Figure \ref{fig:DO_W}). In other words, none of our models connects HS 0704+6153 with SDSS J0301+0508 and SDSS J2239+2259 in a satisfactory way. Given these difficulties, it appears that our current theoretical framework fails to establish a clear evolutionary link between PG 1159 stars and at least some DOZ white dwarfs.

An alternative possibility, raised by \citet{reindl2014b,reindl2014a} on the basis of their spectroscopic analysis of hot white dwarfs and pre-white dwarfs, is that DOZ stars do not belong to the standard PG 1159$-$DO$-$DB$-$DQ evolutionary channel, but instead descend from the so-called O(He) stars. These objects represent a second type of hydrogen-deficient pre-white dwarfs alongside the PG 1159 stars. Their formation is not fully understood, but they are thought to result from low-mass white dwarf mergers \citep{reindl2014a}. Observationally, they differ from PG 1159 stars by their much lower carbon and oxygen abundances, which are in fact similar to those of DOZ white dwarfs ($\log N_{\rm C}/N_{\rm He} \sim -2.5$). Thus, in this scenario, the carbon abundance would remain approximately constant throughout the O(He) and DOZ phases thanks to the wind. In terms of the input parameters of our simulations, this would amount to assuming a low initial carbon mass fraction and a high mass-loss rate, so that the theoretical curve would pass through the DOZ data points roughly horizontally in the $\log N_{\rm C}/N_{\rm He}-\Teff$ diagram. Although this hypothesis effectively removes the need to invoke a DO population with wildly different mass-loss rates, the fact remains that the coolest carbon-polluted objects require a very high mass-loss rate. Given that the wind is likely metal driven and apparently fades below $\Teff \sim 75,000$ K in PG 1159 stars, it is highly doubtful that it persists down to $\Teff \sim 45,000$ K in the much less carbon-rich DOZ white dwarfs. Another problem is that the number counts of O(He) and DOZ stars exhibiting traces of carbon are largely inconsistent. The region of the temperature-gravity plane where O(He) stars are found ($80,000 \ {\rm K} \le \Teff \le 200,000$ K, $5.00 \le \log g \le 6.75$) contains a total of 27 confirmed hydrogen-deficient objects, among which only six are carbon-bearing O(He) stars (see Figure 3 of \citealt{reindl2014b} and references therein). In contrast, on the white dwarf cooling sequence, DOZ stars account for more than half of the hydrogen-deficient population above $\Teff = 60,000$ K (Paper I). Clearly, these objects cannot all descend from O(He) stars.

All things considered, we are still unable to draw a definitive conclusion regarding the origin of DOZ white dwarfs. That said, we can identify a number of avenues to improve the situation. First, it is possible that our ability to interpret the observations is hampered by our simplified treatment of the stellar wind, which is by far the most uncertain feature in our modeling of the PG 1159-to-DO transition. In this context, advances in our theoretical understanding of the source of mass loss in hot white dwarfs are greatly needed. Furthermore, the wind might not be the only transport mechanism acting against the downward diffusion of carbon: the process of radiative levitation might also play a role. More specifically, we can imagine that the wind dominates at very high temperature but then dies out (perhaps around $\Teff \sim 75,000$ K), after which radiative levitation takes over and continues to provide some support against gravitational settling. As a result, the amount of carbon in the outer layers remains non-zero, such that PG 1159 stars naturally evolve into DOZ white dwarfs for a long period of time (possibly down to $\Teff \sim 45,000$ K). The main issue with this scenario is that the surface carbon abundances predicted by existing theoretical calculations of radiative levitation are too low by one and a half order of magnitude ($-4.5 \lta \log N_{\rm C}/N_{\rm He} \lta -3.5$; \citealt{chayer1995}). However, these results were obtained under the assumption of a local equilibrium between gravitational and radiative forces; it is entirely plausible that such an equilibrium has not yet been reached in hot white dwarfs. Therefore, we believe that future chemical evolution simulations should include a time-dependent, non-equilibrium treatment of radiative levitation.

A clearer picture of carbon pollution in hot helium-rich white dwarfs may also be achieved through improvements on the observational front. First, the effective temperatures, stellar masses, and carbon abundances derived from optical spectroscopic analyses have large uncertainties, typically in the range $5000-20,000$ K, $0.05-0.25 \ \msun$, and $0.2-0.5$ dex, respectively \citep{werner2014a,reindl2014b}\footnote{The error on the effective temperature actually becomes smaller for cooler objects. For instance, the two coolest DOZ white dwarfs, SDSS J0301+0508 and SDSS J2239+2259, have uncertainties of $1000-2000$ K in $\Teff$ \citep{reindl2014b}. For this reason, the existence of DOZ stars down to $\Teff \sim 45,000$ K is a robust result.}. It is possible that these uncertainties preclude us from discerning the temperature-mass relation expected from our models for DOZ white dwarfs. Ultraviolet spectroscopy could help improve this situation, as it usually provides more precise and accurate atmospheric parameters for these objects \citep{werner2018}. Second, the total number of analyzed objects is currently relatively low. Hopefully, the ongoing spectroscopic follow-up of Gaia white dwarf candidates \citep{kilic2020,tremblay2020} will reveal new DOZ stars and thereby increase the density of data points in the $\log N_{\rm C}/N_{\rm He}-\Teff$ diagram.

\subsection{The Physical Properties of DQ White Dwarfs} \label{sec:disc_DQ_1}

The carbon abundance pattern of cool DQ white dwarfs is better understood than that of their hot DOZ counterparts. Moreover, their temperature, mass, and carbon abundance estimates are usually much more precise, with typical uncertainties of $\sim$200 K, $\sim$0.05 $\msun$, and $\sim$0.1 dex, respectively \citep{coutu2019}. Thus, the comparison between our various models and the observations should allow us to make meaningful inferences about the physical properties of DQ stars. Nevertheless, we are faced with two major difficulties in the interpretation of the empirical DQ sequence. 

The first difficulty is of theoretical nature and is obvious from Figures \ref{fig:DQ_M} to \ref{fig:DQ_OV}: the parameter space is highly degenerate. Indeed, the predicted evolution of the surface carbon abundance in the DQ phase is affected in a very similar way by the stellar mass, the thickness of the envelope, the initial carbon content, and the extent of convective overshoot. As such, the effect of a change in one of these parameters can be canceled out by varying the other parameters. This means that our standard set of parameter values ($M = 0.6 \ \msun$, $\log q_{\rm env} = -2.0$, $X_{\rm C,i} = 0.5$, and $f_{\rm ov} = 0.075$) is not the only one that provides a good match to the observed DQ sequence. For instance, a smaller amount of overshoot can be compensated by either a lower mass, a thinner envelope, a higher initial carbon abundance, or any combination thereof (see Figures \ref{fig:DQ_M} to \ref{fig:DQ_OV}). Fortunately, some of these parameters are subject to additional constraints. First, it is a well-known fact that the mass distribution of white dwarfs is very sharply peaked at $M \sim 0.6 \ \msun$ \citep{genest-beaulieu2019a,tremblay2019,kilic2020}. Second, pre-white dwarf evolutionary calculations predict that PG 1159 stars should have $\log q_{\rm env} \sim -2.0$ \citep{miller-bertolami2006,althaus2009b}\footnote{In these calculations, the predicted value of $\log q_{\rm env}$ actually varies between $-2.5$ and $-1.5$ depending on the stellar mass, but we argued in Section \ref{sec:res_DQ} that the exact value is of no consequence for the chemical evolution of DQ white dwarfs if $\log q_{\rm env} \gta -2.5$.}. Third, spectroscopic studies show that young PG 1159 stars (in which diffusion has not yet modified the surface composition) typically have $X_{\rm C}/X_{\rm He} \sim 1.2 - 1.5$ \citep{werner2006}. Therefore, our assumptions on the total mass, the envelope mass, and the initial composition rest on solid grounds (although we take a closer look at the $M = 0.6 \ \msun$ hypothesis in the next subsection). On the other hand, the efficiency of overshoot in DQ stars is poorly known, but since it is the sole remaining unconstrained property, it can be probed by calibrating $f_{\rm ov}$ on the observations. This approach effectively provides a new way to investigate convective overshoot in cool white dwarfs. Still, we want to emphasize that our ``best-fitting'' value of the overshoot parameter, $f_{\rm ov} = 0.075$, is only approximate, as it depends on our assumptions on $M$, $q_{\rm env}$, and $X_{\rm C,i}$, as well as on the possible presence of trace hydrogen (see Figure \ref{fig:DQ_H}). In addition, it may change following improvements in the input physics of our models, especially regarding the outer boundary condition: we currently assume a grey atmosphere, while \citet{camisassa2017} showed that a non-grey atmosphere results in a deeper convection zone (and thus likely requires less overshoot)\footnote{According to Figure 9 of \citet{camisassa2017}, the effect of replacing an Eddington atmosphere with a detailed non-grey atmosphere is quite large: the convection zone becomes half an order of magnitude deeper in terms of fractional mass at low effective temperature. However, the effect of upgrading to a detailed non-grey atmosphere is expected to be much smaller in our models, because we assume a grey atmosphere, which is a better approximation than an Eddington atmosphere.}.

The second difficulty is of observational nature. In Figures \ref{fig:DQ_D} to \ref{fig:DQ_OV}, a dotted black line indicates the optical detection limit of carbon for a signal-to-noise ratio ${\rm S/N} = 20$ (which is the median value for the DQ sample of \citealt{coutu2019}), kindly computed by S. Blouin using his DQ model atmospheres \citep{blouin2019a}\footnote{The limit is not very sensitive to the assumed S/N. In particular, the theoretical limit, corresponding to an infinite S/N, is only $\sim$0.3 dex lower than the limit for ${\rm S/N} = 20$.}. The curve closely follows the base of the observed DQ sequence, which implies that the apparent narrowness of the sequence is probably a visibility effect rather than a true feature of the DQ population. In other words, the very tight distribution of DQ stars in the $\log N_{\rm C}/N_{\rm He}-\Teff$ diagram cannot be interpreted as evidence that all carbon-polluted white dwarfs share the exact same mass and chemical structure. Rather, carbon-polluted white dwarfs are likely characterized by a small yet finite range of $M$ and $X_{\rm C,i}$ values, but only the most carbon-polluted ones appear as DQ stars, whereas the others are classified as DC stars because their atmospheric carbon remains optically undetectable. This view is confirmed by the fact that some optically featureless objects do exhibit carbon features in the ultraviolet \citep{weidemann1995}, although their actual carbon content is highly uncertain \citep{dufour2011}. Furthermore, the idea of a range of masses and progenitor abundances is more in line with model-atmosphere analyses of PG 1159, DO, DB and DQ stars \citep{werner2006,reindl2014b,genest-beaulieu2019b,coutu2019} and also with our discussion of the carbon abundance pattern of DOZ white dwarfs in Section \ref{sec:disc_DO}. Therefore, although our calculations show that all white dwarfs descending from PG 1159 stars will inevitably experience the convective dredge-up of settling carbon at low effective temperature, only a fraction of them will actually be identified as DQ white dwarfs based on optical spectroscopy. This fraction cannot be determined at the present time, as this would require knowledge of the true width of the entire carbon abundance distribution, including both DQ and DC stars. However, we can infer from our results that the objects that do become DQ white dwarfs are probably those with lower masses, more carbon-rich progenitors, and/or thinner envelopes. This has an important consequence on our assessment of the extent of convective overshoot: our ``best-fitting'' value of the overshoot parameter, $f_{\rm ov} = 0.075$, might be overestimated, since it is based on the most carbon-polluted members of the white dwarf population originating from PG 1159 stars. As such, this value should be regarded as an upper limit.

In spite of all these challenges, one robust conclusion that can be drawn from our study is that convective overshoot in cool DQ and DC white dwarfs is somewhat less significant than expected from recent hydrodynamical simulations ($f_{\rm ov} \sim 0.2-0.4$; \citealt{kupka2018,cunningham2019}). While this result may seem puzzling, it is in fact perfectly reasonable upon further thought. Hydrodynamical simulations of convective overshoot have so far been performed only for warm DA white dwarfs ($\Teff = 11,000 - 18,000$ K). These objects have shallow, non-adiabatic convection zones extending no deeper than $\log q \sim -12.0$. On the other hand, cool DQ and DC white dwarfs have deep, adiabatic convection zones reaching down to $\log q \sim -5.0$ (see Figure \ref{fig:evol}). As pointed out by \citet{herwig2000} in the context of AGB stars, the efficiency of overshoot is expected to be different in these two regimes: under adiabatic conditions, the convective flows have a higher velocity, and thus when they penetrate the radiative region, they have less time to exchange heat with the surroundings, which causes them to decelerate more quickly. Therefore, as a white dwarf cools and its convection zone deepens and becomes increasingly adiabatic, we do expect the relative extent of its overshoot region (as measured by $f_{\rm ov}$) to decrease. This behavior is partially borne out by the coolest hydrodynamical simulations currently available (T. Cunningham 2021, private communication). 

Our finding has an important implication for the study of cool metal-polluted white dwarfs. These objects owe their surface composition to the accretion of planetary material, and thus the measured heavy-element abundances can be used to deduce the accretion rate and the composition of the accreted matter. This calculation involves the amount of mass contained within the convectively mixed region, concisely called the mixed mass, as well as the diffusion timescales of the relevant metals at the base of this region \citep{dupuis1993,koester2009}. The mixed mass is usually evaluated directly from the mixing-length theory, without considering convective overshoot. \citet{cunningham2019} demonstrated that the inclusion of overshoot (as predicted by their hydrodynamical simulations) leads to a large increase of the inferred accretion rates of warm hydrogen-rich white dwarfs. A similar effect is also expected for helium-rich white dwarfs but has never been assessed due to the unavailability of overshoot calculations for these objects. Our semi-empirical constraint on the overshoot parameter provides for the first time a quantitative estimate of the true mixed mass in cool helium-rich white dwarfs ($\Teff \lta 10,000$ K), which is essential for an accurate determination of the accretion rates of DZ stars. In our coolest models with $f_{\rm ov} = 0.075$, the boundary of the fully mixed region is located $\sim$0.8 pressure scale height below the convection zone. Therefore, when evaluating DZ accretion rates, we recommend to compute the mixed mass and diffusion timescales assuming an overshoot region extending at most 0.8 pressure scale height below the convection zone predicted by the mixing-length theory. We stress that this prescription should actually provide an upper limit on the accretion rates, given that our estimate of the overshoot parameter is itself an upper limit. Besides, we note that metal-accreting white dwarfs can be subject to additional mixing associated with the process of thermohaline convection \citep{deal2013,wachlin2017,bauer2018,bauer2019}. However, this mixing mechanism was found to be negligible in helium-dominated envelopes \citep{deal2013,bauer2019}, hence it does not affect our interpretation.

So far, we have restricted our discussion to the white dwarfs forming the well-defined DQ sequence, but other more unusual objects also deserve a few comments. There are a few cool DQ stars that lie well above the sequence (and thereby coincidentally appear to blend with the massive DQ stars originating from mergers). Our evolutionary calculations suggest that these objects can be explained in three possible ways: low masses, carbon-rich progenitors, and/or thin envelopes. The first possibility is ruled out by the empirical results of \citet{coutu2019} and \citet{blouin2019b}, according to which most of them have typical masses. Furthermore, it is unclear why only a small subgroup of DQ white dwarfs would be characterized by thinner-than-average envelopes. Consequently, the most likely explanation is that they descend from exceptionally carbon-rich PG 1159 stars. Other interesting objects are the two cool DB stars for which \citet{desharnais2008} obtained carbon abundance measurements from ultraviolet spectroscopy and which are shown as asterisks in Figure \ref{fig:full}. Although we cannot conclude anything about these objects because of their metal accretion history, they still highlight the fact that ultraviolet spectroscopy of cool DB white dwarfs could be a valuable tool to further our understanding of the carbon dredge-up phenomenon. This is also true for cooler DC white dwarfs, according to the above discussion.

\subsection{The Observed Mass Distribution of DQ White Dwarfs} \label{sec:disc_DQ_2}

\begin{figure*}
\centering
\includegraphics[width=2.\columnwidth,clip=true,trim=2.0cm 9.25cm 1.25cm 10.5cm]{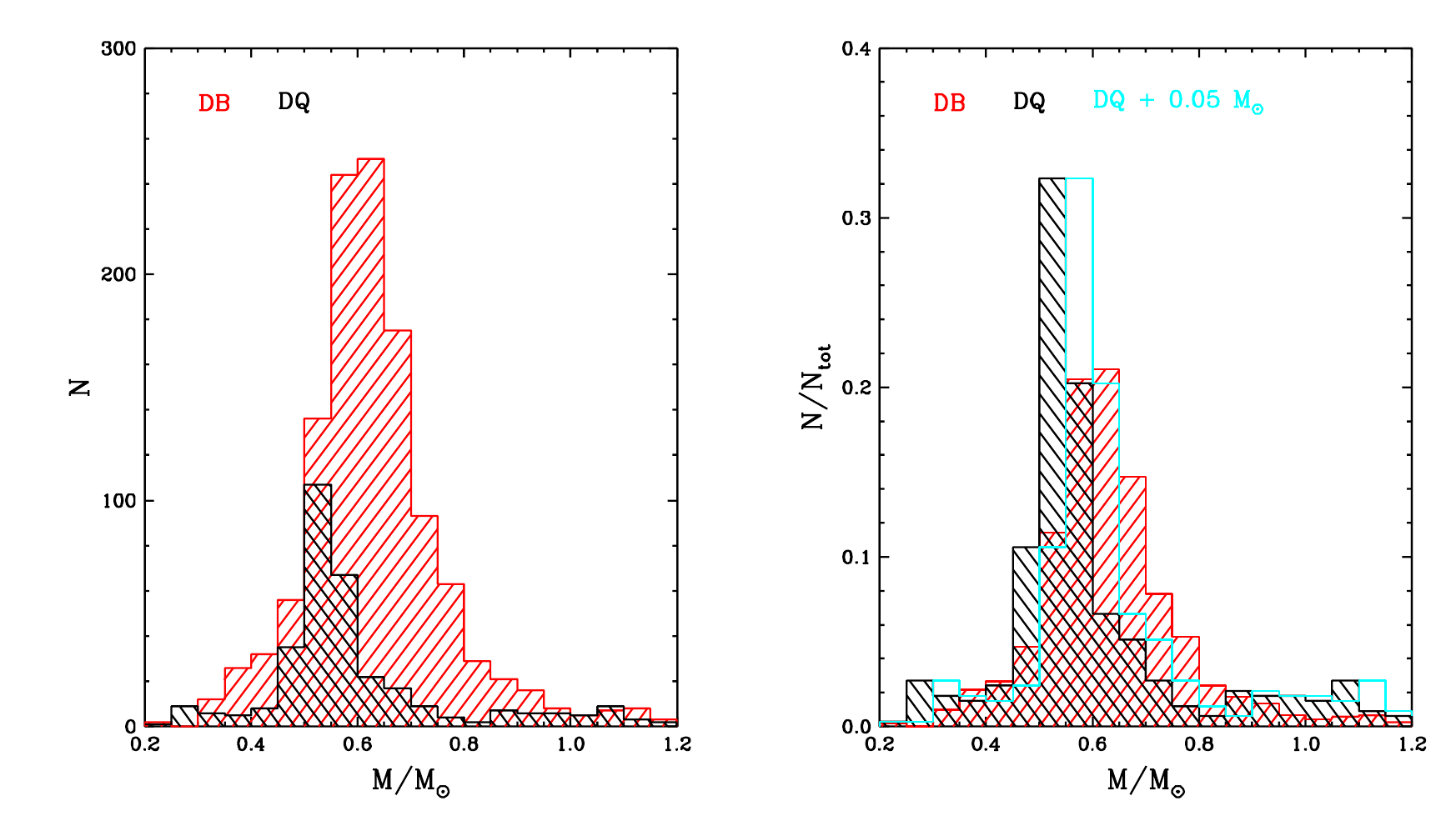}
\caption{Left panel: absolute mass distributions of DB white dwarfs from \citet{genest-beaulieu2019b} and of DQ white dwarfs from \citet{coutu2019} and \citet{blouin2019b}, shown as red and black histograms, respectively. In both cases, we use the so-called photometric masses and exclude all objects with parallax measurement errors larger than 25\%. Right panel: same as the left panel, but for the relative mass distributions. Also displayed as a cyan histogram is the DQ distribution shifted by $+0.05 \ \msun$.}
\vspace{2mm}
\label{fig:mass}
\end{figure*}

In recent years, the advent of the Gaia mission has revolutionized the empirical determination of white dwarf masses. In particular, \citet{coutu2019} relied on new trigonometric parallax measurements from Gaia to obtain for the first time the detailed mass distribution of a large sample of DQ stars (303 objects, or an almost 20-fold increase with respect to the previous analysis of \citealt{dufour2005}). Surprisingly, they found that the mass distribution of DQ white dwarfs peaks at $M \sim 0.55 \ \msun$ and thus appears shifted by $-0.05 \ \msun$ relative to that of their DB precursors. This difference may seem small but is in fact significant given the high precision of the Gaia data. \citet{koester2019} independently obtained the same result using an almost identical dataset but a distinct model-atmosphere grid. The offset is illustrated in Figure \ref{fig:mass}, which compares the mass distributions of DB stars from \citet{genest-beaulieu2019b} and of DQ stars from \citet{coutu2019} and \citet{blouin2019b}, both in absolute (left panel) and relative (right panel) terms. \citet{coutu2019} argued that the systematic shift is likely not real and is possibly caused by the inadequate treatment of ultraviolet opacities in current DQ model atmospheres. They pointed out that this idea is supported by the case of the well-known DQ white dwarf Procyon B, for which their model-derived mass of $0.554 \pm 0.013 \ \msun$ is slightly lower than the precise dynamical mass of $0.592 \pm 0.006 \ \msun$ obtained by \citet{bond2015}.

Our chemical evolution simulations offer a new perspective on this apparent problem. First, as discussed in the previous subsection, DQ stars presumably represent a particularly carbon-rich subset of the whole carbon-polluted white dwarf population. Second, our calculations shown in Figure \ref{fig:DQ_M} predict that the surface carbon abundance is very sensitive to the stellar mass, in the sense that carbon enrichment is more significant in lower-mass models. The obvious conclusion is that trace carbon is preferentially detected in low-mass objects and therefore that DQ white dwarfs are naturally expected to have lower-than-average masses. For this reason, it appears plausible that the mass shift observed by \citet{coutu2019} and \citet{koester2019} is real after all. Figure \ref{fig:mass} seems to support this interpretation: in the left panel, the DQ mass distribution (black histogram) fits the low-mass part of the DB mass distribution (red histogram) relatively well, as anticipated if low-mass and high-mass DB stars generally evolve into DQ and DC stars, respectively, at low effective temperature. That said, one must be careful when comparing the absolute distributions, since the DB and DQ samples were constructed using slightly different selection criteria (although they are both largely based on the sample of spectroscopically confirmed white dwarfs from the Sloan Digital Sky Survey). The relative distributions displayed in the right panel allow a fairer comparison; we can see that DQ white dwarfs (black histogram) are significantly over-represented in the range $0.45 \ \msun \lta M \lta 0.55 \ \msun$ and under-represented in the range $0.60 \ \msun \lta M \lta 0.80 \ \msun$ relative to their DB progenitors (red histogram), which is consistent with the above considerations\footnote{Another difference between the two mass distributions is the presence of a small bump around $M \sim 1.0 \ \msun$ in the DQ distribution. This bump contains the so-called hot and warm DQ stars resulting from white dwarf mergers \citep{coutu2019} and is thus irrelevant to the present discussion.}. Besides, we note that if most DQ stars truly have $M \sim 0.55 \ \msun$, then assuming $M = 0.6 \ \msun$ in our simulations probably led us to overestimate the amount of convective overshoot needed to reproduce the observed carbon abundance pattern (see Section \ref{sec:disc_DQ_1}). Thus, we reiterate that our estimate of the overshoot parameter, $f_{\rm ov} = 0.075$, actually provides an upper limit to the extent of the overshoot region.

Of course, we cannot entirely rule out that issues in DQ model atmospheres are responsible for at least part of the systematic offset. As mentioned above, the case of Procyon B does point in that direction. Nevertheless, even if we artificially increase the masses of all DQ stars by $0.05 \ \msun$, as illustrated in the right panel of Figure \ref{fig:mass} (cyan histogram), our remarks remain valid: the relative mass distributions of DB and DQ stars have markedly different shapes, with the DQ distribution showing both a clear excess at $M \sim 0.55 \ \msun$ and a clear deficit in the range $0.65 \ \msun \lta M \lta 0.80 \ \msun$. We interpret this as the natural consequence of the more efficient carbon dredge-up in lower-mass white dwarfs.

\subsection{Tales of Spectral Evolution} \label{sec:disc_evol}

The evolutionary simulations presented in Paper II and in this paper have allowed us to identify and analyze a number of possible scenarios of white dwarf spectral evolution. One conclusion that stands out is that the sequence of events and the final outcome are primarily dictated by the initial envelope composition and thus by the type of progenitor star. There is of course the trivial case of white dwarfs that inherit a standard ``thick'' hydrogen layer and thereby retain their DA character throughout their life. Although this case is thought to apply to roughly 75\% of the white dwarf population (Paper I), it is of little interest for the present work given that it does not involve any spectral transformation. On the other hand, the chemical diversity of helium-dominated stellar remnants gives rise to an array of spectral evolution channels, which we now better understand thanks to our calculations. At this point, we feel that it is appropriate to give an overview of the emerging picture. The different evolutionary paths can be divided into four categories, according to the presence or absence of hydrogen and carbon in the envelope of the progenitor. 

1. The simplest case is that of an almost pure-helium object, containing virtually no hydrogen and only traces of heavy elements. These characteristics are observed in some O(He) stars \citep{reindl2014a}. With such an initial composition, the formation of a superficial hydrogen layer through diffusion obviously cannot take place, and the dredge-up of settling carbon by the helium convection zone is expected to be insignificant. Consequently, the star always preserves its essentially pure-helium atmosphere as it cools down and successively becomes a DO, DB, and DC white dwarf.

2. A more interesting evolution occurs if the progenitor possesses a small amount of residual hydrogen, as seen in some other O(He) stars \citep{reindl2014a}. In this second scenario, which we modeled in detail in Section 3.2 of Paper II, the white dwarf experiences two major transformations. First, at high temperature, the hydrogen diffuses upward and accumulates at the surface, thereby turning the DO star into a DA star. Then, at low temperature, the thin hydrogen layer becomes convective and mixes with the underlying helium reservoir. As a result, the hydrogen is thoroughly diluted into the helium-rich envelope, and the spectral type accordingly changes from DA to DC. The effective temperature at which this mixing episode occurs depends on the total hydrogen content: the thicker the hydrogen layer, the lower the mixing temperature \citep{rolland2018,cunningham2019}. Furthermore, note that the atmosphere of a DC white dwarf produced by this evolutionary channel contains an invisible yet non-zero amount of hydrogen, the exact abundance also being determined by the total hydrogen content.

3. The third possibility is a progenitor with significant quantities of carbon and oxygen, namely, a PG 1159 star \citep{werner2006}. The details of the ensuing chemical evolution were extensively studied in the present work: in essence, the sedimentation of carbon and oxygen produces a pure-helium atmosphere, which remains so until the convection zone brings a small amount of carbon back to the surface. As we have seen, the outcome is either a DQ or a DC white dwarf, depending on whether the final carbon abundance is above or below the optical detection limit (as determined by the stellar mass, the thickness of the envelope, the initial carbon abundance, and the extent of convective overshoot). As in the previous case, a DC star formed in this way does not have a pure-helium atmosphere, but this time the contaminant is carbon instead of hydrogen.

4. Finally, the fourth possible type of progenitor is a combination of the two previous ones: a PG 1159 star with trace hydrogen. The spectral evolution of such an object, which was investigated in Section \ref{sec:res_H} of the present paper, combines all of the above-mentioned phenomena. At high temperature, the carbon and oxygen sink out of sight while the hydrogen floats to the surface, thus giving rise to a quick PG 1159-to-DO-to-DA transition. At low temperature, the presence of a thin hydrogen layer prevents carbon dredge-up at first, but once the convective mixing of the hydrogen and helium layers occurs, the convection zone very suddenly expands inward and reaches into the carbon reservoir. Because the hydrogen is ultimately diluted and thus ``hidden'' in the helium-dominated envelope, the DA white dwarf becomes either a DQ or a DC white dwarf, once again depending on the magnitude of carbon enrichment. In the end, this evolutionary channel produces a helium-rich atmosphere polluted by both hydrogen and carbon. Besides, we note that the hydrogen/helium mixing event and the concomitant onset of carbon dredge-up are expected to be further delayed in objects with higher hydrogen contents. As an extreme example, in the so-called hybrid PG 1159 stars (which show hydrogen lines in addition to the usual helium, carbon, and oxygen lines), the surface hydrogen mass fraction is typically $X_{\rm H} \sim 0.1$ \citep{werner2006}; assuming a homogeneous composition down to $\log q \sim -4.0$ \citep{althaus2005b}, this corresponds to a total hydrogen mass of $\sim$10$^{-5} M$. Once all this hydrogen has floated up, the superficial hydrogen layer will be so thick that it will never mix with the underlying helium layer \citep{rolland2018}, such that the DA-to-DQ/DC transition will never take place. Therefore, hybrid PG 1159 stars will remain DA white dwarfs at low effective temperature.

All these considerations have a particularly interesting implication concerning the surface composition of DC and DZ white dwarfs. Through a comprehensive model-atmosphere analysis relying on Gaia parallax measurements, \citet{bergeron2019} found that the assumption of a pure-helium atmosphere leads to a global overestimation of the masses of DC stars by $\sim$0.1 $\msun$. \citet{coutu2019} reported a similar problem for DZ stars (which differ from their DC counterparts only by their metal accretion history). In both cases, it was argued that more reasonable masses could only be obtained by including an undetectable amount of hydrogen or carbon in the model atmospheres (see \citealt{bergeron2019} and \citealt{coutu2019} for details). The results of our simulations, combined with our knowledge of pre-white dwarf objects, are perfectly consistent with this empirical requirement. Indeed, the majority of the pre-white dwarf population consists of carbon-rich PG 1159 stars (see Figure 3 of \citealt{reindl2014b} and references therein). Moreover, most O(He) stars, when observed at sufficiently high signal-to-noise ratio, reveal traces of hydrogen \citep{reindl2014a}. In other words, there exist very few (if any) white dwarf progenitors with virtually pure-helium envelopes, which suggests that the first spectral evolution channel described above is extremely uncommon. In this context, our calculations predict that nearly all DC and DZ white dwarfs should have atmospheres containing invisible amounts of either hydrogen, carbon, or both. Our work therefore lends strong theoretical support to the results of \citet{bergeron2019} and \citet{coutu2019} regarding the surface composition of cool helium-rich white dwarfs.

\section{Conclusions} \label{sec:conclu}

In this paper, we presented an in-depth theoretical investigation of the PG 1159$-$DO$-$DB$-$DQ evolutionary channel based on state-of-the-art simulations of element transport in evolving white dwarfs. We examined the influence of several stellar parameters and physical processes on the variation of the atmospheric carbon abundance over a very large range of effective temperature. Our study represents a significant improvement over previous works, as it is the only one to date that combines a sophisticated treatment of element transport, an extensive exploration of the parameter space, and a good agreement with observations of cool DQ stars. Our main conclusions can be summarized as follows:

1. At high effective temperature, thermal diffusion must be considered in order to model the PG 1159-to-DO transformation properly. The inclusion of this process accelerates the sedimentation of carbon and oxygen and thereby leads to an earlier PG 1159-to-DO transition, by $\sim$3000 K in terms of temperature.

2. The effective temperature at which a PG 1159 star turns into a DO white dwarf primarily depends on the stellar mass and the wind mass-loss rate, in the sense that the spectral change takes place at lower temperature if the mass is lower and/or the wind is stronger. In particular, for our standard mass-loss law (Equation \ref{eq:mdot}), the transition temperature is $\Teff \sim$ 66,000, 75,000 and 83,000 K for $M \sim$ 0.5, 0.6, and 0.7 $\msun$, respectively.

3. The observed carbon abundance pattern of DOZ white dwarfs appears mostly consistent with the idea that these objects represent an intermediate evolutionary stage between the PG 1159 and DO stars, provided that the PG 1159 population is characterized by a reasonable range of masses and mass-loss rates. Nevertheless, there are a few problems with this interpretation, the most serious being that the existence of relatively cool DOZ white dwarfs ($\Teff \sim 45,000 - 50,000$ K) cannot be explained satisfactorily within our current theoretical framework. Radiative levitation, which we did not take into account, might play a more important role than expected.

4. At low effective temperature, thermal diffusion, non-ideal diffusion, and convective overshoot all have a significant impact on the chemical evolution. Thermal diffusion and non-ideal diffusion both enhance the rate of carbon settling and thereby reduce the magnitude of carbon dredge-up, while convective overshoot makes the homogeneously mixed region deeper and thus has the opposite effect.

5. The surface carbon abundance in the DB and DQ phases is highly sensitive to a number of physical parameters, namely, the stellar mass, the thickness of the envelope, the initial carbon content, and the extent of convective overshoot. More specifically, a lower mass, a thinner envelope, a higher initial carbon abundance, and a more efficient overshoot all result in a more important carbon pollution. In our standard model, which assumes $M = 0.6 \ \msun$, $\log q_{\rm env} = -2.0$, $X_{\rm C,i} = 0.5$, and $f_{\rm ov} = 0.075$, the carbon abundance reaches a maximum at $\Teff \sim 10,000$ K and then decreases, hence closely matching the empirical DQ sequence.

6. The chemical evolution is also substantially altered if a small amount of residual hydrogen is present in the PG 1159 progenitor. In this case, the hydrogen initially diffuses upward and accumulates at the surface. This thin hydrogen layer first delays the onset of carbon dredge-up, but then mixes with the underlying helium layer, which leads to a rapid expansion of the convection zone and therefore a sudden carbon enrichment of the outer envelope. The carbon abundance subsequently decreases, as in the hydrogen-free case.

7. The comparison between the predicted and observed composition of DQ stars allowed us to constrain for the first time the extent of the overshoot region in cool helium-rich white dwarfs ($\Teff \lta 10,000$ K). Our inferred upper limit on the overshoot parameter, $f_{\rm ov} \le 0.075$, implies that the overshoot region extends at most 0.8 pressure scale height below the convection zone predicted by the mixing-length theory. This result should be taken into account in the determination of metal accretion rates of DZ stars.

8. Our calculations offer an elegant explanation for the difference between the empirically derived mass distributions of DB and DQ white dwarfs: although all DB stars likely experience carbon dredge-up, this phenomenon is more significant (and thus more easily detected) in lower-mass objects, which translates into a relative excess and deficit of low-mass and high-mass DQ stars, respectively.

9. Given the fact that the vast majority of helium-rich pre-white dwarf objects show traces of hydrogen and/or carbon, our simulations predict that the atmospheres of nearly all cool helium-rich white dwarfs should contain small (potentially invisible) amounts of hydrogen and/or carbon. This result is perfectly in line with recent model-atmosphere analyses of DC and DZ stars.

10. There are many ways by which our understanding of the PG 1159$-$DO$-$DB$-$DQ evolutionary channel could be further improved in the future. At the hot end, more theoretical work on the source of winds and on the effects of radiative levitation would be highly desirable. Moreover, the identification and analysis of new DOZ stars in the sample of Gaia white dwarf candidates would be beneficial as well. At the cool end, the most promising avenue to obtain a more accurate picture of the carbon-bearing white dwarf population is definitely to investigate the atmospheric carbon content of DB and DC stars through ultraviolet spectroscopy.

Finally, we want to point out that our knowledge of carbon pollution in helium-rich white dwarfs is still far from complete. For instance, the chemical evolution of the very massive DQ stars remains unclear. Although the emergence of the hot DQ white dwarfs at $\Teff \sim 25,000$ K can be explained by the convective dredge-up of carbon in extremely thin helium envelopes ($\log q_{\rm env} \sim -8.0$; \citealt{althaus2009a}), no model so far has been able to reproduce the carbon abundance pattern of their alleged progeny, the warm DQ white dwarfs \citep{brassard2007,coutu2019}. Furthermore, the presence of trace carbon at the surface of some hot DB stars ($\Teff > 20,000$ K) is even more mysterious. One hypothesis that has been proposed is that a residual stellar wind slows down the gravitational settling of carbon down to $\Teff \sim 20,000$ K \citep{fontaine2005}; however, there is currently no physical basis for the occurrence of such a cool wind. In fact, no known mechanism provides a satisfactory explanation to the phenomenon \citep{koester2014}. Therefore, it appears that the theory of the spectral evolution of white dwarfs still faces many fascinating challenges.

\acknowledgments

We are grateful to Simon Blouin for a careful reading of our manuscript and for providing us with the optical detection limit of carbon based on his DQ model atmospheres. We also thank Tim Cunningham for helpful discussions regarding convective overshoot in hydrodynamical simulations. This work was supported by the Natural Sciences and Engineering Research Council (NSERC) of Canada and the Fonds de Recherche du Qu\'ebec $-$ Nature et Technologie (FRQNT).

\bibliographystyle{aasjournal}
\bibliography{main}

\end{document}